\documentclass[final]{siamonline171218}

\usepackage{amsmath,amssymb}

\usepackage{lipsum}
\usepackage{amsfonts}
\usepackage{graphicx}
\usepackage{epstopdf}

\usepackage{algorithmic}
\ifpdf
  \DeclareGraphicsExtensions{.eps,.pdf,.png,.jpg}
\else
  \DeclareGraphicsExtensions{.eps}
\fi

\newcommand{\TheTitle}{High dimensional inference for the structural health monitoring of lock gates.} 
\newcommand{\TheAuthors}{Matthew Parno, Devin O'Connor, and Matthew Smith}

\headers{High dimensional inference for SHM of lock gates.}{\TheAuthors}

\title{{\TheTitle}}

\author{
  Matthew Parno\thanks{US Army Cold Regions Research and Engineering Laboratory, Hanover, NH
    (\email{Matthew.D.Parno@usace.army.mil}).}
  \and
  Devin O'Connor \thanks{US Army Cold Regions Research and Engineering Laboratory, Hanover, NH
    (\email{Devin.T.O'Connor@usace.army.mil}).}
  \and
  Matthew Smith\thanks{US Army Coastal and Hydraulics Laboratory, Vicksburg, MS
    (\email{Matthew.D.Smith@usace.army.mil}).}
}

\usepackage{amsopn}

\usepackage{bm}
\usepackage{subcaption}
\usepackage{siunitx}

\usepackage{tikz}
\usetikzlibrary{calc,patterns}

\usepackage[utf8]{inputenc}
\usepackage{pgfplots}
\usepgfplotslibrary{groupplots}
\pgfplotsset{compat=1.12}
\DeclareMathOperator{\Tr}{Tr}

\newcommand{\strain}{\varepsilon}
\newcommand{\strainvec}{\bm{\strain}}

\newcommand{\stress}{\sigma}

\newcommand{\displ}{u}
\newcommand{\displvec}{\bm{\displ}}

\newcommand{\hydro}{f}
\newcommand{\hydrovec}{\bm{\hydro}}

\newcommand{\quoin}{w_q}
\newcommand{\quoinvec}{\bm{w}_q}

\newcommand{\miter}{w_m}
\newcommand{\mitervec}{\bm{w}_m}

\newcommand{\ddim}{{N_{d}}}

\ifpdf
\hypersetup{
  pdftitle={\TheTitle},
  pdfauthor={\TheAuthors}
}
\fi

\begin{document}

\maketitle

\begin{abstract}
Locks and dams are critical pieces of inland waterways.  However, many components of existing locks have been in operation past their designed lifetime.  To ensure safe and cost effective operations, it is therefore important to monitor the structural health of locks.  To support lock gate monitoring, this work considers a high dimensional Bayesian inference problem that combines noisy real time strain observations with a detailed finite element model.  To solve this problem, we develop a new technique that combines Karhunen-Lo\`eve decompositions, stochastic differential equation representations of Gaussian processes, and Kalman smoothing that scales linearly with the number of observations and could be used for near real-time monitoring.  We use quasi-periodic Gaussian processes to model thermal influences on the strain and infer spatially distributed boundary conditions in the model, which are also characterized with Gaussian process prior distributions.  The power of this approach is demonstrated on a small synthetic example and then with real observations of Mississippi River Lock 27, which is located near St. Louis, MO USA.  The results show that our approach is able to probabilistically characterize the posterior distribution over nearly 1.4 million parameters in under an hour on a standard desktop computer.
\end{abstract}

\begin{keywords}
  Structural health monitoring, Bayesian inference, Gaussian processes, Model error, Uncertainty quantification, Parameter reduction
 \end{keywords}

\begin{AMS}
60G15, 62F15, 65C20
\end{AMS}

\section{Introduction}\label{sec:intro}
Inland waterways, especially the Mississippi river system, provide an important transportation network within the United States.  Indeed, inland waterways operated by the U.S. Army Corps of Engineers (USACE) delivered more than \$229 billion worth of  cargo in 2015 \cite{ASCEReportCard2017}.  This is only possible because of an extensive system of locks and dams that make these waterways navigable.  However, much of this critical infrastructure has been in operation past its designed service life and is at risk of excessive fatigue and failure.  With increasing age, unscheduled maintenance and downtime also become significant concerns, especially in light of constrained maintenance budgets.  It is estimated that an unscheduled outage of Lock 27 on the Mississippi River has an economic impact of nearly \$2.8 million per day \cite{Gillerman2013}.  Approaches for characterizing the state of a gate, estimating remaining life, and minimizing the cost of necessary maintenance are therefore critical for safe and cost-effective operation.  

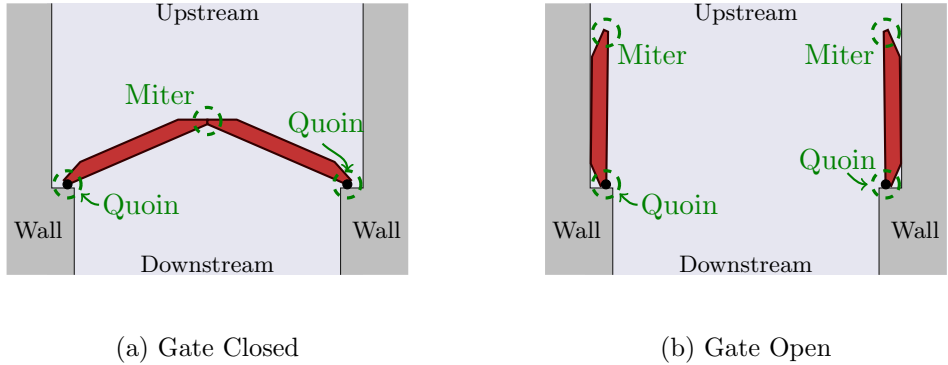
\begin{figure}[h!]
\centering
\begin{subfigure}{0.45\textwidth}
\centering
\begin{tikzpicture}[scale=0.6]

\fill[fill=blue!40!black!10!white] (-1,4) rectangle (6,-2);

\begin{scope}[xshift=-0.55cm]

\draw[fill=black!25,draw=none] (-1.35,4) -- (-0.35,4) -- (-0.35,-0.08) -- (0.15,-0.08) -- (0.15,-2) -- (-1.35,-2) -- cycle;
\draw[draw=black!90] (-0.35,4) -- (-0.35,-0.08) -- (0.15,-0.08) -- (0.15,-2);

\draw[fill=red!70!black!80!white, draw=red!20!black, thick]  (-0.080000,0.095000) -- (0.282500,0.495000) -- (2.450000,1.430000) -- (3.100000,1.430000) -- (3.100000,1.330000) -- (0.040000,0.010000) -- cycle;
\draw[fill=black] (0.0,0.0) circle (0.1cm);
\end{scope}

\begin{scope}[xshift=5.65cm, xscale=-1]

\draw[fill=black!25,draw=none] (-1.35,4) -- (-0.35,4) -- (-0.35,-0.08) -- (0.15,-0.08) -- (0.15,-2) -- (-1.35,-2) -- cycle;
\draw[draw=black!90] (-0.35,4) -- (-0.35,-0.08) -- (0.15,-0.08) -- (0.15,-2);

\draw[fill=red!70!black!80!white, draw=red!20!black, thick]  (-0.080000,0.095000) -- (0.282500,0.495000) -- (2.450000,1.430000) -- (3.100000,1.430000) -- (3.100000,1.330000) -- (0.040000,0.010000) -- cycle;
\draw[fill=black] (0.0,0.0) circle (0.1cm);
\end{scope}

\node at (-1.2,-1) {\footnotesize Wall};s
\node at (6.3,-1) {\footnotesize Wall};

\node at (2.55,3.75) {\footnotesize Upstream};
\node at (2.55,-1.75) {\footnotesize Downstream};

%

\draw[dashed, very thick, draw=green!50!black] (2.55,1.4) circle (0.3cm);
\node[anchor=south east, green!50!black] at (2.5,1.55) {Miter};

\draw[dashed, very thick, draw=green!50!black] (-0.55,0.0) circle (0.3cm);
\node[anchor=north west, green!50!black] (quoin1) at (0.0,0.0) {Quoin};
\draw[thick,green!50!black] (0.15,-0.45) edge[out=180,in=-45,->] (-.25,-0.2);

\draw[dashed, very thick, draw=green!50!black] (5.65,0.0)  circle (0.3cm);
\node[anchor=south west, green!50!black] (quoinlab) at (4.1, 0.8) {Quoin};
\draw[thick,green!50!black] (5.1,0.9) edge[->,out=-45,in=90] (5.65,0.3);

\end{tikzpicture} 
\caption{Gate Closed}
\end{subfigure}
\begin{subfigure}{0.45\textwidth}
\centering
\begin{tikzpicture}[scale=0.6]

\fill[fill=blue!40!black!10!white] (-1,4) rectangle (7,-2);

\begin{scope}[xshift=-0.55cm]

\draw[fill=black!25,draw=none] (-1.35,4) -- (-0.35,4) -- (-0.35,-0.08) -- (0.15,-0.08) -- (0.15,-2) -- (-1.35,-2) -- cycle;
\draw[draw=black!90] (-0.35,4) -- (-0.35,-0.08) -- (0.15,-0.08) -- (0.15,-2);

\begin{scope}[rotate=66]
\draw[fill=red!70!black!80!white, draw=red!20!black, thick]  (-0.080000,0.095000) -- (0.282500,0.495000) -- (2.450000,1.430000) -- (3.100000,1.430000) -- (3.100000,1.330000) -- (0.040000,0.010000) -- cycle;
\draw[fill=black] (0.0,0.0) circle (0.1cm);
\end{scope}
\end{scope}

\begin{scope}[xshift=5.65cm, xscale=-1]

\draw[fill=black!25,draw=none] (-1.35,4) -- (-0.35,4) -- (-0.35,-0.08) -- (0.15,-0.08) -- (0.15,-2) -- (-1.35,-2) -- cycle;
\draw[draw=black!90] (-0.35,4) -- (-0.35,-0.08) -- (0.15,-0.08) -- (0.15,-2);

\begin{scope}[rotate=66]
\draw[fill=red!70!black!80!white, draw=red!20!black, thick]  (-0.080000,0.095000) -- (0.282500,0.495000) -- (2.450000,1.430000) -- (3.100000,1.430000) -- (3.100000,1.330000) -- (0.040000,0.010000) -- cycle;
\draw[fill=black] (0.0,0.0) circle (0.1cm);
\end{scope}

\end{scope}

\node at (-1.2,-1) {\footnotesize Wall};
\node at (6.3,-1) {\footnotesize Wall};

\node at (2.55,3.75) {\footnotesize Upstream};
\node at (2.55,-1.75) {\footnotesize Downstream};

%

\draw[dashed, very thick, draw=green!50!black] (-0.55,0.0) circle (0.3cm);
\node[anchor=north west, green!50!black] (quoin1) at (0.0,0.0) {Quoin};
\draw[thick,green!50!black] (0.15,-0.45) edge[out=180,in=-45,->] (-.25,-0.2);

\draw[dashed, very thick, draw=green!50!black] (-0.55, 3.35)  circle (0.3cm) node[green!50!black,below right]{Miter};

\draw[dashed, very thick, draw=green!50!black] (5.65, 3.35)  circle (0.3cm) node[green!50!black, below left]{Miter};

\draw[dashed, very thick, draw=green!50!black] (5.65,0.0) circle (0.3cm);
\node[anchor=south west, green!50!black] (quoinlab) at (3.5, 0.0) {Quoin};
\draw[thick,green!50!black] (4.8,0.25) edge[->,out=-45,in=180] (5.25,0.0);

\end{tikzpicture}
\caption{Gate Open}
\end{subfigure}

\caption{Plan view of miter gate.  This style of lock gate is formed from two leaves that press together and seal when the upstream water level is higher than the downstream level.  The edge where the leaves connect is called the miter and the edge of a leaf that touches the wall is call the quoin.  A lock is composed of two miter gates, an upstream gate and downstream gate, but only one is shown here.}
\label{fig:mitergeom}
\end{figure}

Miter lock gates, illustrated in Figure \ref{fig:mitergeom}, are the most common type of lock gate in the USACE portfolio and require expensive maintenance to  avoid failures like those at the Markland Lock and Dam \cite{Chapman2010} and the Chickamauga Lock and Dam \cite{smith2017concrete}.   Previous efforts for monitoring these gates, namely \cite{Eick2018}, have focused on damage detection and cannot identify the location or cause of the damage.  To overcome this, we   employ a detailed finite element model of a particular USACE lock gate, Lock 27, and leverage recent developments in the Bayesian statistics and uncertainty quantification communities to infer unknown model boundary conditions.  The data for our approach are strain observations from the USACE SMART Gate database \cite{smartgate}.  Working these SMART Gate data pose many real-world issues that are not commonly considered in the uncertainty quantification community.  Temporally correlated observations, biased sensors, and incredibly large parameter spaces, in particular, prevent the direct application of standard Bayesian techniques.  To overcome this, we develop a new statistical strategy that employs a novel combination of Karhunen-Lo\`eve decompositions and statespace representations of Gaussian processes.  We introduce this approach within the scope of our miter gate problem, but it is generally applicable in other areas as well.

The use of a detailed finite element model allows us to accurate characterize the stress field throughout the miter gate and thus identify locations of potential damage or excessive fatigue.  However, accurate predictions of the stress state require an accurate representation of the gate's geometry and knowledge of the model's boundary conditions.  Unfortunately, boundary conditions for the model (i.e., external loads), and their dependence on time and water level, are generally not known.  The boundary conditions are dependent on the contact between the gate and wall, which is spatially variable and can be compromised by corrosion and the formation of gaps.  Because much of the gate is underwater, it also difficult to measure the contact directly.

To obtain the stress state and enable structural health monitoring, we therefore need to use available strain observations to characterize the boundary loads as a function of time, location on the gate, and water level.    To accomplish this, we adopt a Bayesian approach and define a statistical model for the strain observations by combining the finite element model with Gaussian process priors and likelihoods.  A statistical approach is required because the observations are noisy and cannot the inverse problem is ill posed; the observations alone cannot completely constrain the high dimensional boundary conditions.  A Bayesian approach allows us to overcome these challenges while also enabling further uncertainty quantification in predictions of fatigue and remaining life.

Despite their incredible power, Bayesian approaches in structural health monitoring have have seen relatively limited use on high dimensional problems with complex forward models.  One reason for this is computational expense.  Even inverse problems with linear models and Gaussian distributions, which admit analytic solutions, can become computationally intractable for standard approaches as the number of observations (e.g., $>10000$) or parameter dimension grows.  This is partly because the number of floating point operations required to directly evaluate a Gaussian density grows cubically with the number of observations.  Fortunately, recent connections between Gaussian processes (GP) and stochastic differential equations (SDEs) can help alleviate this cubic growth.  Below, we will develop a novel approach that combines model reduction via static condensation, parameter reduction via Karhunen-Lo\`eve (KL) decompositions, the SDE representation of Gaussian processes, and Kalman smoothing.  The basic idea is to separate stochastic, spatial, and temporal dependencies with a KL decomposition, transform Gaussian process priors into SDEs, and then apply a standard Kalman smoothing approach.

In Section \ref{sec:background}, we provide a limited review of miter gates and Bayesian methods in structural health monitoring.   Section \ref{sec:forwardmodel} describes our linear elastic forward model and degree of freedom reduction technique. Our inverse problem formulation is presented in Section \ref{sec:formulation}, where we discuss our prior, error, and likelihood formulations. A illustrative smaller scale example to test our proposed methods is discussed in Section \ref{sec:toy} and the solution procedure and results of the inverse approach to the full scale lock gate is presented in Section \ref{sec:solution} and Section \ref{sec:results}. Lastly, some conclusions are drawn from our study in Section \ref{sec:conclusions}.

\section{Background}\label{sec:background}

\subsection{Miter Gates}
Locks are a critical component of inland waterways that allow boats and barges to pass around dams and rapids by acting as hydraulic elevators.  Within a lock system, there are multiple gates that work in concert to block the flow of water and raise or lower water inside the lock chamber.  Miter gates are a particularly prevalent type of lock gate with two leaves that press together to form a seal.  A single miter gate is illustrated in Figure \ref{fig:mitergeom} in both a closed configuration, which allows the water levels to change, and an open configuration, which allows water traffic to enter or leave the gate.

When working properly, a miter gate works like a three pin arch to transmit large hydrostatic loads to the walls.   However, corrosion can cause gaps to form between the quoin (see Figure \ref{fig:mitergeom}) and wall, increasing stresses in the gate.  This extra stress can result in  increased fatigue and premature failure.  It is therefore important to monitor quoin-wall forces and quickly identify gaps or other potential issues that could prevent safe operation.

\subsection{Bayesian Structural Health Monitoring}
Bayesian methods have been used in structural health monitoring for nearly two decades.   A general Bayesian structural health monitoring framework was first presented in the seminal work of \cite{beck1998updating}.   The authors used a Laplace approximation at each local maximum of the posterior to create a Gaussian mixture approximation of the posterior.  Since then, Laplace approximations have seen extensive use in Bayesian structural health monitoring \cite{Vanik2000, yuen2006unified}.   However, as \cite{ching2007transitional} points out, Laplace approximations have limited utility when the data is uninformative and posterior is skewed or far from Gaussian.   Furthermore, when vibrational (i.e., modal) observations are employed, the posterior often contains multiple modes and can be difficult to explore.  Markov chain Monte Carlo (MCMC) algorithms have been developed in \cite{beck2002bayesian} and \cite{ching2007transitional} to overcome this challenge, but are limited to low dimensional settings.   Relatively low dimensional Bayesian inference problems have also been used more recently for identifying parameters in civil infrastructure models \cite{Astroza2017,Ebrahimian2017} and optimally placing sensors \cite{Flynn2010}.

In general, previous Bayesian structural health monitoring efforts focus on the use of modal data to infer a few lumped parameters.  In that setting, a fundamental challenge is exploring multimodal, but low dimensional, posterior distribution.    On the other hand, we employ a limited number of strain observations and infer high dimensional spatially and temporally distributed parameter fields.  Our model is linear and our posterior is Gaussian, which means we can analytically derive the posterior mean and covariance.  However, unlike previous structural health monitoring efforts, our parameter space is incredible large.  This makes it intractable to directly compute the posterior and poses challenges that are fundamentally different than those tackled in previous structural health monitoring work.  Our work is also different than the recent work of \cite{Eick2018}, which focused on detecting damage to lock gates, but not identifying possible reasons for the damage.

\subsection{Bayesian Inference}
Let $\bm{w}_r$ denote unknown model parameters (e.g., boundary loads) and let $\hat{\strainvec}_{\text{obs}}$ denote a vector of observational data.   Bayesian approaches describe the unknowns $\bm{w}_r$ and the observations $\strainvec_{\text{obs}}$ as random variables.  We assume that the actual observations $\hat{\strainvec}_{\text{obs}}$ are a realization of  $\strainvec_{\text{obs}}$.    The distribution of the parameters $\bm{w}_r$ represents our degree of belief that $\bm{w}_r$ takes certain values.  The goal of Bayesian inference is therefore to characterize the posterior density $\pi(\bm{w}_r | \strainvec_{\text{obs}} = \hat{\strainvec}_{\text{obs}})$.\footnote{Notice that by discussing Bayes' rule in terms of densities, we have implicitly assumed that the distribution of $\bm{w}_r$ admits a probability density.}  Bayes' rule is then given by
\begin{equation}
\pi(\bm{w}_r | \strainvec_{\text{obs}} = \hat{\strainvec}_{\text{obs}}) = \frac{ \pi(\strainvec_{\text{obs}} = \hat{\strainvec}_{\text{obs}}| \bm{w}_r ) \pi(\bm{w}_r )} {\int \pi(\strainvec_{\text{obs}} = \hat{\strainvec}_{\text{obs}}| \bm{w}_r ) \pi(\bm{w}_r ) d \bm{w}_r},
\end{equation}
where $\pi(\strainvec_{\text{obs}} = \hat{\strainvec}_{\text{obs}}| \bm{w}_r )$ is the likelihood function, and $\pi(\bm{w}_r )$ is the prior density.  

The prior density incorporates any information known about the parameters $\bm{w}_r$ before considering the observations $\hat{\strainvec}_{\text{obs}}$.  Examples include: knowledge about the sign of $\bm{w}_r$, knowledge about correlations between the components of $\bm{w}_r$, or beliefs about reasonable mean values.   

The likelihood function is essentially a statistical model (perhaps with physical components) of the data.  It captures what we would expect the observation random variable $\strainvec_{\text{obs}}$ to look like if the parameters $\bm{w}_r$ were known.  A common way to construct the likelihood is to simply introduce additive noise to an existing physical model.  For example, let $\strainvec_{\text{obs}}$ be strain at several strain gages and assume we have a physically-based structural model $F(\bm{w}_r)$ that predicts strain at these locations.   A common practice is then to introduce a zero-mean additive error term $\bm{e}\sim N(0, \Sigma_{e})$ such that
\begin{equation}
\strainvec_{\text{obs}} = F(\bm{w}_r) + e.
\end{equation}
With this form, the likelihood function is simply a multivariate Gaussian density with mean $F(\bm{w}_r)$ and covariance $\Sigma_e$.   While simple and computationally convenient, zero mean additive Gaussian errors are rarely accurate models of the observation random variables.  Missing physics and numerical discretizations are common examples of error sources that would result in more systematic differences between model predictions and real-world observations that are not well characterized by iid Gaussian errors.   Section \ref{sec:formulation} will describe more sophisticated alternatives using correlated Gaussian processes to more accurately model the observations in our miter gate problem.

\subsection{Gaussian Processes}\label{sec:background:gp}
Gaussian processes are the extension of multivariate Gaussian distributions to function spaces.    Unlike Gaussian distributions, which are completely defined by a mean vector and covariance matrix, Gaussian processes are completely defined by a mean function and covariance kernel.  Let $\theta$ denote the input to the Gaussian process (e.g., position).   We will use the notation $GP(\mu(\theta), k(\theta, \theta^\prime))$ to denote a Gaussian process with mean function $\mu(\theta)$ and covariance kernel $k(\theta, \theta^\prime)$.  There are many canonical covariance kernels (see \cite{Rasmussen2006} for a more comprehensive list), but our focus will be on various combinations of Matern, squared exponential, periodic, and white noise kernels.   Matern kernels, denoted by $k_\nu(\theta, \theta^\prime)$, take the form
\begin{equation}
k_\nu(\theta, \theta^\prime) = \sigma^2 \frac{2^{1-\nu}}{\Gamma(\nu)}\left( \sqrt{2\nu} \frac{\|\theta - \theta^\prime \|_2}{L}\right)^\nu K_v\left(\sqrt{2\nu}\frac{\|\theta - \theta^\prime \|_2}{L}\right), \label{eq:matern_kernel}
\end{equation}
where $\sigma^2$ is the variance of the process, $\Gamma$ is the gamma function, $K_\nu$ is the modified Bessel function of the second kind, $\nu$ is a smoothness parameter governing the differentiability of the process, and $L$ is a lengthscale parameter.

Taking the limit as $\nu\rightarrow \infty$, the Matern kernel converges to the squared exponential kernel and has the form
\begin{equation}
k_{se}(\theta, \theta^\prime) = \sigma^2 \exp\left( -\frac{(\theta - \theta^\prime)^2}{2L^2} \right).
\end{equation}
This kernel is sometimes referred to as the radial basis function (RBF) kernel and results in a Gaussian process whose realizations are infinitely differentiable.

Notice that $k_\nu$ an $k_{se}$ decrease monotonically as the $\theta$ and $\theta^\prime$ increases, which is not always necessary.   For example, the standard periodic kernel described in \cite{Rasmussen2006} has peaks of equal height and equal spacing, which results in a periodic Gaussian processes.   This kernel takes the form
\begin{equation}
k_p(\theta, \theta^\prime) = \sigma^2 \exp\left[-\frac{2}{L^2} \sin^2\left(\frac{\pi}{P}\|\theta-\theta^\prime\|\right)\right], \label{eq:periodic_kernel}
\end{equation}
where $\sigma^2$ is the variance of the process, $L$ is a lengthscale, and $P$ is the period of the correlation.  Typically, $k_p$ is multiplied by another aperiodic kernel like $k_\nu$ to create a quasi-periodic kernel.    Realizations of a quasi-periodic are locally periodic, but can change over time.  This is useful for modeling time series that have regular cycles (e.g., diurnal cycles in air temperature) but are not perfectly periodic.

The white noise covariance kernel, denoted here by $k_\delta(\theta, \theta^\prime)$, takes the form
\begin{equation}
k_\delta(\theta, \theta^\prime) = \sigma^2 \delta(\theta,\theta^\prime),
\end{equation}
where $\sigma^2$ is again the variance of the process and $\delta(\theta, \theta^\prime)$ is the Dirac delta function.

Notice that all of these kernels are for scalar-valued Gaussian processes and can only be applied directly when modeling a single dependent variable.   However, we will be interested in vector-valued Gaussian processes.  For example, we will model thermal strain as a vector-valued Gaussian process with several dependent variables; one variable for each strain gage.    The advantage of using a single vector-valued process over several independent scalar-valued processes is that the vector-valued Gaussian process is able to capture correlation between the dependent variables, like the correlation in thermal strain that results from similar air temperature fluctuations at different strain gages.

As reviewed in \cite{Alvarez2012}, there are many ways to extend scalar covariance kernels to describe vector-valued processes.  We will employ the linear model of coregionalization, which is common in many fields \cite{Rasmussen2006, Alvarez2012, Wackernagel2013} and often referred to as co-kriging in the geostatistics community.  The idea is to represent the correlated components of the vector-valued process as linear combinations of independent Gaussian processes.   Let $\bm{\strain}(\theta)$ be the vector-valued process with $N$ components and let $z_1(\theta),z_2(\theta),\ldots,z_M(\theta)$ be $M$ independent Gaussian processes.  With the linear model of coregionalization, $\bm{\strain}(\theta)$ is represented as
\begin{equation}
\bm{\strain}(\theta) = A \left[ \begin{array}{c} z_1(\theta) \\ \vdots \\ z_M(\theta) \end{array}\right],
\end{equation}
for some $M\times M$ matrix $A$.  If each $z_i$ has a marginal variance of $1$, then the covariance of $\bm{\strain}(\theta)$ is given by $AA^T$.  Thus, if the marginal covariance of $\bm{\strain}(\theta)$ is known, $A$ can be constructed from the matrix square root of the covariance.



\section{Forward model}\label{sec:forwardmodel}

The general structure of a miter gate is shown in Figure \ref{fig:mitergeom}.   Two such gates are necessary to form a chamber where the water level can be raised and lowered, working as an elevator for boats and barges.   We will only consider a single leaf of the gate and will be interested in the conditions at the Quoin-Wall interface and the Miter-Miter interface. 

\subsection{Linear elasticity}
Momentarily ignoring boundary conditions, the stress in the gate satisfies
\begin{equation}
\nabla \cdot \stress(x) = F(x) \quad \forall\, x \in \Omega, \label{eq:conservation}
\end{equation}
where $\stress(x)$ is the stress tensor at point $x$, $F(x)$ is a vector of external loads, and $\Omega$ is the interior domain of the miter gate.  In this work, we assume that the material density is constant and that no other body forces are present.

We employ a linear-elastic constitutive model of the form
\begin{eqnarray}
\stress(x) & = & C : \strain_e(x) \\
&=& 2\mu\strain_e(x) + \lambda \Tr{\left(\strain_e(x)\right)} I,
\end{eqnarray}
where $C$ is the fourth order elasticity tensor defined by Lam\`e parameters $\mu$ and $\lambda$.   We use small strain theory to relate the displacement $\displ$ and strain $\strain_e$ with
\begin{equation}
\strain_e(x) = \frac{1}{2}\left[\nabla \displ(x) + \nabla \displ(x)^T\right].
\end{equation}
Small strain theory is applicable here because the displacement is quite small compared to the size of the gate.

Stresses are prescribed at the Quoin-Wall and Miter-Miter boundaries, as well as a hydrostatic force from the upstream and downstream water levels.  Let $h^{+}$ denote the upstream water height, $h^{-}$ denote the downstream water height, and $x$ denote position.  The prescribed boundary loads then take the form of the Neumann boundary conditions
\begin{eqnarray}
\stress(x) \cdot \hat{n}(x) &=& \quoin(x, h^{+},h^{-},t) \quad \forall \,x \in \Gamma_Q. \label{eq:qbc}\\
\stress(x) \cdot\hat{n}(x) &=&\miter(x, h^{+},h^{-},t) \quad \forall \,x \in \Gamma_M \label{eq:mbc}\\
\stress(x) \cdot\hat{n}(x) &=& \hydro(x,h^{+},h^{-})  \quad \forall \, x \in \Gamma_N. \label{eq:wbc},
\label{eq:neumannbc}
\end{eqnarray}
where $\quoin(x,h^{+},h^{-},t)$ is the load on the quoin-wall boundary at time $t$, $\miter(x,h^{+},h^{-},t)$ is the load on the miter-miter interface, and $\hydro(x,h^{+},h^{-},t)$ represents the hydrostatic forces on both the upstream and downstream faces of the gate.  

We assume the gate geometry does not change significantly over time, which would violate our small-strain assumption, and thus $\hydro(x,h^{+}, h^{-})$ does not have an explicit dependence on time.  This is in contrast to the Quoin-Wall and Miter-Miter boundary loads $\quoin(x, h^{+},h^{-},t)$ and $\miter(x, h^{+},h^{-},t)$, which may change as a result of corrosion, fatigue, crack formation, or other time-varying processes.  Thus, we model $\quoin(x, h^{+},h^{-},t)$ and $\miter(x, h^{+},h^{-},t)$ with an explicit dependence on $t$.

The miter gate geometry is quite complex and prevents the analytic solution of \eqref{eq:conservation}--\eqref{eq:neumannbc}.   We therefore adopt a finite element method to discretize \eqref{eq:conservation}--\eqref{eq:neumannbc} and form a linear system.  Using boldface to denote the discretized variables, the discretized analog of \eqref{eq:conservation} becomes 
\begin{eqnarray}
\bm{K} \displvec &=& \hydrovec(h^{+},h^{-}) + \quoinvec(h^{+},h^{-},t) + \mitervec(h^{+},h^{-},t), \label{eq:basesys}\\
\strainvec_e &=& \bm{B} \displvec \label{eq:strainmat} 
\end{eqnarray}
where $\bm{K}$ is the stiffness matrix and $\bm{B}$ is a differentiation matrix mapping displacement to strain.  Notice that $\quoinvec$ and $\mitervec$ are vectors containing weights on the finite element basis functions and no longer depend on location $x$.

We know the boundary loads $\quoinvec(h^{+},h^{-},t)$ and $\mitervec(h^{+},h^{-},t)$ depend on water level and time.  However, we do not know the form of these relationships.    Our goal is therefore to characterize the functions $\quoinvec(h^{+},h^{-},t)$ and $\mitervec(h^{+},h^{-},t)$ using the finite element model defined by \eqref{eq:basesys}--\eqref{eq:strainmat} and observations of the strain field $\strain_e$ at several locations in the model domain $\Omega$.

\subsection{Static condensation}\label{sec:model:cond}

The displacement vector $\displvec$ contains degrees of freedom across the entire gate.  However, for inference we are only interested in degrees of freedom along the quoin, the miter, and near the strain gage locations.  We can exploit this fact using static condensation \cite{wilson1974static} to reduce the dimension of $\displvec$ and subsequently accelerate inference.

To accomplish this, we decompose the linear system from \eqref{eq:basesys} into two components
\begin{equation}
\bm{K} = \left[\begin{array}{cc} \bm{K}_{11} & \bm{K}_{12}\\ \bm{K}_{21} & \bm{K}_{22}\end{array}\right] \left[\begin{array}{c}\displvec_c(h^{+},h^{-},t) \\ \displvec_r(h^{+},h^{-},t) \end{array}\right] = \left[\begin{array}{c}\hydrovec_1(h^{+},h^{-}) \\\hydrovec_2(h^{+},h^{-}) + \bm{w}_r(h^{+},h^{-},t)\end{array}\right],
\end{equation}
where $\displvec_r$ denotes our desired ``reduced" degrees of freedom near the boundaries and strain gages,  and $\displvec_c$ denotes the remaining degrees of freedom, the ``complementary" degrees of freedoms.   Notice that the quoin-wall and miter-miter loads in $\bm{w}_r$ are only present in the second component of the right hand side because all boundary degrees of freedom are contained in $\bm{u}_r$.

In our case, the dimension of $\displvec_r$ is much less than the dimension of $\displvec_c$.  We exploit this fact to partially invert the stiffness matrix offline and accelerate subsequent model evaluations.  Performing a partial block Gaussian elimination (e.g., computing the Schur complement) leads to a system for $\displvec_r$ alone
\begin{eqnarray}
\left[\bm{K}_{22} - \bm{K}_{21}\bm{K}_{11}^{-1}\bm{K}_{12}\right]\displvec_r & = & \hydrovec_2 - \bm{K}_{21}\bm{K}_{11}^{-1}\hydrovec_1 + \bm{w}_2\\
\bm{K}_r \displvec_r& = & \hydrovec_r + \bm{w}_r, \label{eq:reduced_displacement}
\end{eqnarray}
where $\bm{K}_r = \bm{K}_{22} - \bm{K}_{21}\bm{K}_{11}^{-1}\bm{K}_{12}$ is the Schur complement of $\bm{K}$ and $\hydrovec_r = \hydrovec_2 - \bm{K}_{21}\bm{K}_{11}^{-1}\hydrovec_1$.  The differentiation matrix $\bm{B}$ can also be reduced to obtain a matrix $\bm{B}_r$ satisfying
\begin{equation}
\strainvec_{er} = \bm{B}_r \displvec_r. \label{eq:reduced_strain}
\end{equation}
Combining  \eqref{eq:reduced_displacement} and \eqref{eq:reduced_strain} yields the reduced model
which implies
\begin{equation}
\strainvec_{er} = \bm{B}_r\bm{K}_r^{-1} \left[\bm{f}_r + \bm{w}_r\right]. \label{eq:redmodel}
\end{equation}
Notice that this expression defines a relationship between the unknown loads $\bm{w}_r$ and strain, which can be observed.   It will therefore play a role in the likelihood function described below.

\section{Inverse Problem Formulation}\label{sec:formulation}
\subsection{Gaussian Process Prior}\label{sec:prior}

Before considering the discretized load vector $\bm{w}_r$ in \eqref{eq:redmodel}, we will formulate the Bayesian inference problem on the continuous load functions $\quoin$ and $\miter$.  To clarify dependencies, we denote all random variables as functions of $\omega$, where $\omega$ is an abstract random variable defined on an probability space $(S, \mathcal{F}, \mathbb{P})$.  We will also use $h=[h^+,h^-]$ to simplify notation. The quoin and miter loads will thus be denoted by $\quoin(x,h,t,\omega)$ and $\miter(x,h,t,\omega)$, respectively.   

We use Gaussian processes to model the prior distributions over the quoin and miter loads, 
\begin{eqnarray}
\quoin(x, h, t, \omega) &\sim& GP\left( \mu_q(x, h), k_q\left([x, h, t],[x^\prime, h^\prime, t^\prime]\right)\right)\\
\miter(x, h, t, \omega) &\sim& GP\left( \mu_m(x, h), k_m\left([x, h, t],[x^\prime, h^\prime, t^\prime]\right)\right),
\end{eqnarray}
for some mean functions $\mu_q$ and $\mu_m$ as well as covariance kernels $k_q$ and $k_m$. The mean functions $\mu_q(x, h)$ and $\mu_m(x, h)$ are obtained by replacing \eqref{eq:qbc}--\eqref{eq:mbc} with Dirichlet conditions $\displ(x) = 0$ on the quoin and miter boundaries and then extracting the loads $\stress(x)\cdot \hat{n}(x)$ along the boundaries.  These mean functions reflect our belief that, when operating properly, there should be perfect contact at the quoin wall and no significant displacement should occur.  

Miter gates transfer hydrostatic loads through horizontal girders into the quoin-wall and miter-miter boundaries.  Thus, the boundary loads near these horizontal girders are typically larger than in between girders.  The loads are these points are also more sensitive to changes in conditions.  To reflect this, we expect the variance of the loads $\quoin(x,h,t,\omega)$ and $\quoin(x,h,t,\omega)$ to be larger near the girders, with smooth transitions between.  To model this behavior, we scale the load covariance by the prior mean.   To see this mathematically, assume the covariance kernels $k_q$ and $k_m$ take a tensor product form 
\begin{eqnarray}
k_q\left([x, h, t],[x^\prime, h^\prime, t^\prime]\right) &=&  k_{qx}(x,x^\prime)k_{qh}(h,h^\prime)k_{qt}(t,t^\prime)\\
k_m\left([x, h, t],[x^\prime, h^\prime, t^\prime]\right) &=& k_{mx}(x,x^\prime)k_{mh}(h,h^\prime)k_{mt}(t,t^\prime).
\end{eqnarray}
We use a squared exponential kernel to characterize the water level dependence in $k_{qh}$ and $k_{mh}$ because we expect the loads to vary smoothly (almost linearly) with water height.   The time kernels $k_{qh}$ and $k_{mh}$ are chosen as Matern kernels for computational reasons that will become clear in Section \ref{sec:solution}.  The position kernels $k_{qx}$ and $k_{mx}$ incorporate the prior mean scaling and are slightly more complicated.   They take the form of a Matern kernel that is weighted by the prior mean.  In particular, $k_{qx}$ and $k_{mx}$ take the form
\begin{eqnarray}
k_{qx}(x,x^\prime) &=& (|\mu_q(x,\bar{h})| + \beta )(|\mu_q(x^\prime,\bar{h})| +\beta) k_{\nu}(x,x^\prime) \label{eq:quoin_tensor}\\
k_{mx}(x,x^\prime) &=& (|\mu_m(x,\bar{h})| + \beta )(|\mu_m(x^\prime,\bar{h})|+\beta) k_{\nu}(x,x^\prime) \label{eq:miter_tensor},
\end{eqnarray}
where $k_{\nu}$ is a Matern kernel.   These nonstationary kernels allow for larger variances near the horizontal girder locations where  $|\mu_q(x,\bar{h})|$ and $|\mu_m(x,\bar{h})|$ are large, but also put a minimum level on the variance, allowing the posterior to capture unexpected behavior in between girders, where $|\mu_q(x,\bar{h})|$ and $|\mu_m(x,\bar{h})|$ are small.

\subsection{Model Error and Likelihood}\label{sec:error}

Assume that strain can be observed at several locations $x_1,x_2,\ldots, x_\ddim$ on the lock gate and let 
\begin{equation}
\hat{\bm{\strain}}_{\text{obs}}(t) = [\hat{\strain}_{\text{obs},1}(t),\hat{\strain}_{\text{obs},2}(t),\ldots,\hat{\strain}_{\text{obs},\ddim}(t)] \label{eq:observedStrain}
\end{equation}
denote the vector of these strains observed at time $t$.  The structural finite element model can be used to make predictions of the elastic strain at the same locations as the observations.     These are denoted by 
\begin{equation}
\bm{\strain}_{er}(t,\omega) = [\strain_{e,1}(t,\omega), \strain_{e,2}(t,\omega), \ldots, \strain_{e,\ddim}(t,\omega)],
\end{equation}
and can be computed from the model with
\begin{equation}
\bm{\strain}_{er}(t,\omega) = \bm{B}_r\bm{K}_r^{-1}\left[\bm{f}_r(h(t)) + \bm{w}_r(h(t), t,\omega)\right] \label{eq:elasticReduced}
\end{equation}

Environmental influences like thermal effects, are not present in the elastic model defined by \eqref{eq:basesys}--\eqref{eq:strainmat}, but can have a significant impact on the observed strain.   The physical strain gages placed on the gate to generate \eqref{eq:observedStrain} also have unknown systematic biases.  Thus, there is a discrepancy between the elastic model and the real-world observations.

Like \cite{Kennedy2001}, we will characterize the model discrepancy with Gaussian processes.  As \cite{Brynjarsdottir2014} points out, incorporating model discrepancy needs to be done with care to avoid a situation where the inference parameters (i.e., boundary loads) and discrepancy are confounded and cannot both be identified from the same data.   Fortunately, we know the general structure of the thermal and bias components, which can help prevent this.

Based on our knowledge of miter gates and strain gages, we model the discrepancy with three terms: the unmodeled thermally-induced strain $\bm{\strain}_T(t,\omega)$, observation bias stemming from strain gage calibration $\bm{\strain}_b(\omega)$, and independent random observation noise $\bm{\epsilon}(t,\omega)$. The prior predictive strain, denoted by $\bm{\strain}_{\text{obs}}(t,\omega)$, is the sum of these components and $\bm{\strain}_{er}$.  Mathematically, we have
\begin{equation}
\bm{\strain}_{\text{obs}}(t,\omega) = \bm{\strain}_{er}(t, \bm{w}_r, \omega)  + \bm{\strain}_T(t, \omega) + \bm{\strain}_b(\omega) + \bm{\epsilon}(t,\omega).
\label{eq:error}
\end{equation}
The actual observations $\hat{\bm{\strain}}_{\text{obs}}(t)$ defined in \eqref{eq:observedStrain} will be treated as a single realization of $\bm{\strain}_{\text{obs}}(t,\omega)$.   

We model each of the components in \eqref{eq:error} as a Gaussian processes with mean functions and covariance kernels that are based on physical intuition for this system.  This careful choice of mean function and covariance kernel will prevent confounding of the model discrepancy (i.e., $\bm{\strain}_T(t, \omega)$, $\bm{\strain}_b(\omega)$, $\bm{\epsilon}(t,\omega)$) with the inference parameters $\bm{w}_r$.  The following sections describe each component in more detail.

\subsubsection{Elastic Strain Gaussian Process}\label{sec:elasticGP}
As a result of using a linear elastic model, we can decompose the elastic strain $\bm{\strain}_e$  into two components: the strain resulting from hydrostatic loads, and the strain resulting from loads on the boundary.   We will denote these by $\strainvec_{e,w}$ and $\strainvec_{e,b}$ respectively, so that 
\begin{equation}
\strainvec_{er}(h,t, \omega) = \strainvec_{er,w}(h) + \strainvec_{er,b}(h,t,\omega).
\end{equation}
Notice that the hydrostatic strain is not a random variable; we have accurate measurements of the water levels $h^+$ and $h^-$ and a good understanding of the hydrostatic loads that they induce.    However, the boundary strain is a random variable because the loads on the boundaries of the lock gate (e.g., Quoin and Miter) are not known exactly and will be inferred.

Using the linear elastic model in \eqref{eq:redmodel}, we obtain
\begin{eqnarray}
\strainvec_{er,w}(h) & = & \bm{B}_r\bm{K}_r^{-1}\bm{f}_r(h)\\
\strainvec_{er,b}(h,t,\omega) & = & \bm{B}_r\bm{K}_r^{-1}\bm{w}_r(h,t,\omega)
\end{eqnarray}
The boundary load vector $\bm{w}_r$ has two components, $\bm{w}_{rQ}$ and $\bm{w}_{rM}$, corresponding to the quoin and miter sides of the gate.  We can therefore write
\begin{equation}
\bm{w}_r(h,t,\omega) = \left[\begin{array}{c} \bm{w}_{rQ}(h,t,\omega) \\ \bm{w}_{rM}(h,t,\omega) \end{array}\right],
\end{equation}
and independently specify the mean and covariance structure of $\bm{w}_{rQ}(h,t,\omega)$ and $\bm{w}_{rM}(h,t,\omega)$ using the prior defined in Section \ref{sec:prior}.

\subsubsection{Thermal Strain Gaussian Process}

Changing air temperatures, water temperatures, and solar radiation can cause the temperature of the of the lock gate itself to vary dramatically.  This causes expansion and contraction of the gate and induces a thermal strain.  The observed strain is a combination of the thermal and elastic strains but the the thermal component is not captured in our structural model.   The $\bm{\strain}_T(t,\omega)$ term in \eqref{eq:error} is included to account for this.  We assume $\bm{\strain}_T(t,\omega)$ is a vector-valued Gaussian process with mean function $\mu(t)$ and adopt the linear model of coregionalization discussed in Section \ref{sec:background:gp}.  Let $\Sigma_T$ denote the marginal covariance of $\bm{\strain}_T(t,\omega)$ and let $\sqrt{\Sigma_T}\in \mathbb{R}^N\times\mathbb{R}^N$ denote a matrix square root satisfying $\sqrt{\Sigma_T}\sqrt{\Sigma_T}^T = \Sigma_T$.  We represent the thermal strain $\bm{\strain}_T(t,\omega)$ as
\begin{equation}
\bm{\strain}_T(t,\omega) = \mu_T(t) + \sqrt{\Sigma_T} \left[ \begin{array}{c} z_{T,1}(t,\omega)\\ \vdots \\ z_{T,N}(t,\omega) \end{array}  \right], \label{eq:thermalForm}
\end{equation}
where the components $z_{T,i}(t,\omega)$ are iid Gaussian processes with mean zero and covariance kernel $\kappa_T(t,t^\prime)$.  To capture the impact of diurnal and seasonal temperature variations on the thermal strain, we set $\kappa_T$ to be a quasi-periodic covariance kernel of the form
\begin{equation}
\kappa_T(t,t^\prime) = \kappa_{T1}(t,t^\prime)\kappa_{T2}(t,t^\prime) + \kappa_{T3}(t,t^\prime),
\label{eq:thermal_gp_kernel}
\end{equation}
where $\kappa_{T1}(t,t^\prime)$ is a periodic kernel with the same form as $k_p$ in \eqref{eq:periodic_kernel}, $\kappa_{T2}(t,t^\prime)$ is a Matern kernel, and $\kappa_{T3}(t,t^\prime)$ is another Matern kernel.   Combined, the product $\kappa_{T1}(t,t^\prime)\kappa_{T2}(t,t^\prime)$ defines a quasi-periodic kernel that captures the diurnal, but not perfectly periodic, features of thermal strain.   Of course, air and water temperatures are not entirely periodic and can contain more slowly varying trends over several days.   The third kernel $\kappa_{T3}(t,t^\prime)$ allows us to capture these non-diurnal trends.

\subsubsection{Gage Bias Distribution}

Most strain gages are constructed with wire or metallic foil formed into a small grid pattern that changes electrical resistance when stretched.  A Wheatstone bridge can then be used to measure the resistance change, and thus the strain.  However, a Wheatstone bridge is constructed from three other resistors that need to be balanced when the strain gage is installed \cite{Hoffmann1989}.    A bridge that is not exactly balanced will result in a constant bias in the strain measurements.  The $\bm{\strain}_b(t)$ term in \eqref{eq:error} models this potential bias as a Gaussian random variable, i.e.,
\begin{equation}
\bm{\strain}_b(\omega) \sim N\left(0, \Sigma_b\right),
\label{eq:gage_bias_gp}
\end{equation}
where $\Sigma_b$ is the covariance between the biases on each strain gage. Note that in this formulation, $\bm{\strain}_b(\omega)$ does not depend on time, but can be represented as a Gaussian process with a constant covariance kernel of the form $k(t,t^\prime) = \sigma^2$.  This fact will be useful in Section \ref{sec:solution}, where we will interpret $\bm{\strain}_b(\omega)$ as the solution of temporal stochastic differential equation.

\subsubsection{Additive noise}
Our final error term $\bm{\epsilon}(t)$ accounts for other random observation noise.  Again we assume a Gaussian distribution for $\bm{\epsilon}(t)$, but here each component of $\bm{\epsilon}(t)$ is iid and we have
\begin{equation}
\bm{\epsilon}(t,\omega)  \sim N\left(0, \sigma_\epsilon^2 I\right),
\end{equation}
where $\sigma_\epsilon^2$ is the variance of the observation noise.

\section{Illustrative Example}\label{sec:toy}
To illustrate our approach without the solution complexity and computational costs of working with the full miter gate model, we will now consider a simpler problem.  In this example, we generate synthetic strain observations by explicitly modeling the thermal-elastic behavior of a cantilever beam.  Figure \ref{fig:beam} shows the geometry of this problem as well as the thermal and mechanical boundary conditions.  Our goal is to infer the traction $w_t(h^+)$ and normal $w_n(h^+)$ loads at the left boundary (e.g.\, $\Gamma_Q$ in Figure \ref{fig:beam}) from noisy strain observations in the center of the beam.

\begin{figure}[h!]
\centering

\begin{subfigure}{0.9\textwidth}
\centering
\begin{tikzpicture}[scale=0.8, every node/.style={scale=0.8}]
\tikzstyle{ground}=[fill,pattern=north east lines,draw=none]

\draw[fill=black!10] (0,0) rectangle (10,2);
\node[anchor=south west] at (0,0) {$\Omega$};
\node[anchor=center] at (5,1) {$\nabla\cdot\stress(x) = -a\left[T(x,t)-T_0\right]$};

\draw[ground] (10,-0.5) rectangle (10.25,2.5);
\draw[thick] (10,-0.5) -- (10,2.5);

\node[anchor=east] at (10,1) {\color{blue!70!black}$\displ(x)=0$};
\draw[very thick,blue!70!black] (10,0) -- (10,2) node[above,anchor=south east] {$\Gamma_D$};

\draw[very thick,red!70!black] (0,0) -- (0,2) node[above,anchor=south] {$\Gamma_Q$};
\node[anchor=east] at (-0.01,1) {\color{red!70!black} $\stress(x)\cdot\hat{n}(x) = \left[\begin{array}{cc} w_n(h^+) \\ w_t(h^{+})\end{array}\right]$};

\draw[thick,draw=green!40!black] (0,0) -- (10,0);
\node[anchor=north] (load) at (5,-0.01) {\color{green!40!black} $\stress(x)\cdot\hat{n}(x)= f(h^+)$};

\end{tikzpicture}
\caption{Structural system.  Dirichlet conditions on the right fix the displacement at $0$.  A distributed load is specified on the bottom and left boundaries.  The bottom boundary mimics the known hydrostatic load from the lock gate and the boundary condition on the left mimics the unknown force acting at the quoin-wall interface.}
\end{subfigure}

\begin{subfigure}{0.9\textwidth}
\centering

\begin{tikzpicture}[scale=0.8, every node/.style={scale=0.8}]
\tikzstyle{ground}=[fill,pattern=north east lines,draw=none]

\draw[fill=black!10] (0,0) rectangle (10,2);
\node[anchor=south west] at (0,0) {$\Omega$};
\node[anchor=center] at (5,1) {$\rho c_p \frac{\partial T}{\partial t} - \nabla \cdot (K\nabla T) = 0$};

\node[anchor=west] at (10,1) {\color{blue!70!black}$\nabla T(x,t) \cdot \hat{n}(x) = 0$};
\draw[very thick,blue!70!black] (10,0) -- (10,2);

\draw[very thick,red!70!black] (0,0) -- (0,2);
\node[anchor=east] at (-0.01,1) {\color{red!70!black} $\nabla T(x,t) \cdot \hat{n}(x) = 0$};

\draw[thick,draw=green!40!black] (0,0) -- (10,0);
\node[anchor=north] (temp1) at (5,-0.01) {\color{green!40!black} $T(x,t)=T_b(t)$};

\draw[thick,draw=green!40!black] (0,2) -- (10,2);
\node[anchor=south] (temp2) at (5,2.01) {\color{green!40!black} $T(x,t)=T_t(t)$};

\end{tikzpicture}
\caption{Thermal system.  The left and right boundaries are insulated.  The top and bottom boundaries are immersed in fluids with temperature $T_t$ and $T_b$, respectively.}
\end{subfigure}

\caption{Toy problem geometry and boundary conditions.}
\label{fig:beam}
\end{figure}

\begin{figure}[h!]
\centering
\includegraphics[width=\textwidth]{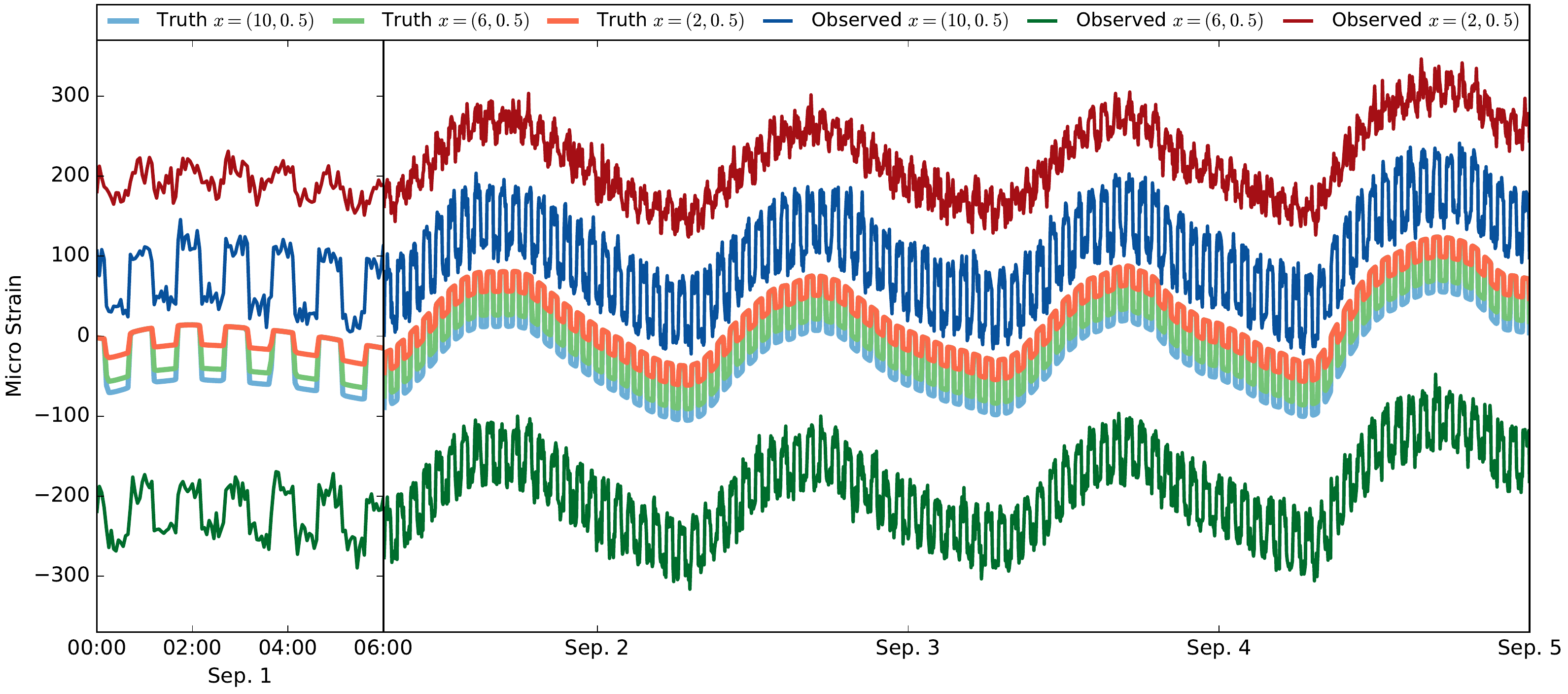}
\caption{Thermally-influenced strain observations for the cantilever beam example.  The first six hours of September 1st are enlarged to illustrate details.  Notice how the observed strains are noisy and offset from the true (i.e., simulated) strains.}
\label{fig:toyObs}
\end{figure}

The beam extends from its lower left corner at $(0,0)$ to its upper right corner at $(12,1)$ and strain is observed at three spatial locations evenly spread across the midline of the beam: $(2,0.5)$, $(6,0.5)$, and $(10,0.5)$.  To generate the synthetic strain observations, the coupled thermal-elastic system from Figure \ref{fig:beam} is simulated over time with prescribed temperatures for $T_a(t)$ and $T_b(t)$ and an oscillating square wave water level $h^+(t)$.  The temperatures are taken from meteorological observations at Lambert-St. Louis international airport in St. Louis, MO from September 1, 2016 to September 5, 2016, which is close to the Lock 27 gate considered in Section \ref{sec:results}.  Noise stemming from both strain gage calibration biases $\bm{\strain}_b$ and random observation errors $\bm{\epsilon}(t)$ is then added  to the simulated strain to create our synthetic observations.  The bias we use is $\bm{\strain}_b = \left[\num{2e-4}, \num{-2e-4}, \num{1e-4}\right]^T$ and the noise $\bm{\epsilon}(t)$ is randomly drawn white noise with variance $\num{1e-10}$ and zero mean. Figure \ref{fig:toyObs} shows the thermally-influenced synthetic strains with and without adding noise.

Clearly, there is a daily cycle in the strains caused by the boundary conditions $T_a(t)$ and $T_b(t)$.  We model these cycles using the quasi-periodic thermal strain covariance kernel in \eqref{eq:thermal_gp_kernel}.  
We set the marginal thermal covariance $\Sigma_T$ to  
\begin{equation}
\Sigma_T = \left[ \begin{array}{ccc} \sigma_{1}^2 & \rho \sigma_{1}\sigma_{2} & \rho \sigma_{1}\sigma_{3}\\ \rho \sigma_{1}\sigma_{2} & \sigma_{2}^2 & \rho \sigma_{2}\sigma_{3} \\ \rho \sigma_{1}\sigma_{3} & \rho \sigma_{2}\sigma_{3} & \sigma_{3}^2 \end{array}\right],
\end{equation}
where $\sigma_{1}^2=\sigma_{2}^2=\sigma_{3}^2=5\times 10^{-9}$ and $\rho=0.9$.  In this example, we use a kernel with the same form as \eqref{eq:thermal_gp_kernel}.  The $k_{T1}$ component is a Matern kernel with $\nu=5/2$, lengthscale $L=0.8$ days, and $\sigma^2=0.5$.  The $k_{T2}$ component is a periodic kernel $k_p$ with period $P=1$ days, lengthscale $L=0.8$, and variance $\sigma^2=1$.   The final $k_{T3}$ kernel is also a Matern kernel with smoothness parameter $\nu=3/2$, lengthscale $L=5$ days, and variance $\sigma^2=0.5$.

The sensor bias $\bm{\varepsilon}_b$ is distributed according to a mean zero Gaussian with covariance $10^{-2}I$, where $I$ is the identity matrix.  The observation noise is a white noise process with mean zero and variance $10^{-10}$.  

\begin{figure}[h!]
\centering
\includegraphics[width=\textwidth]{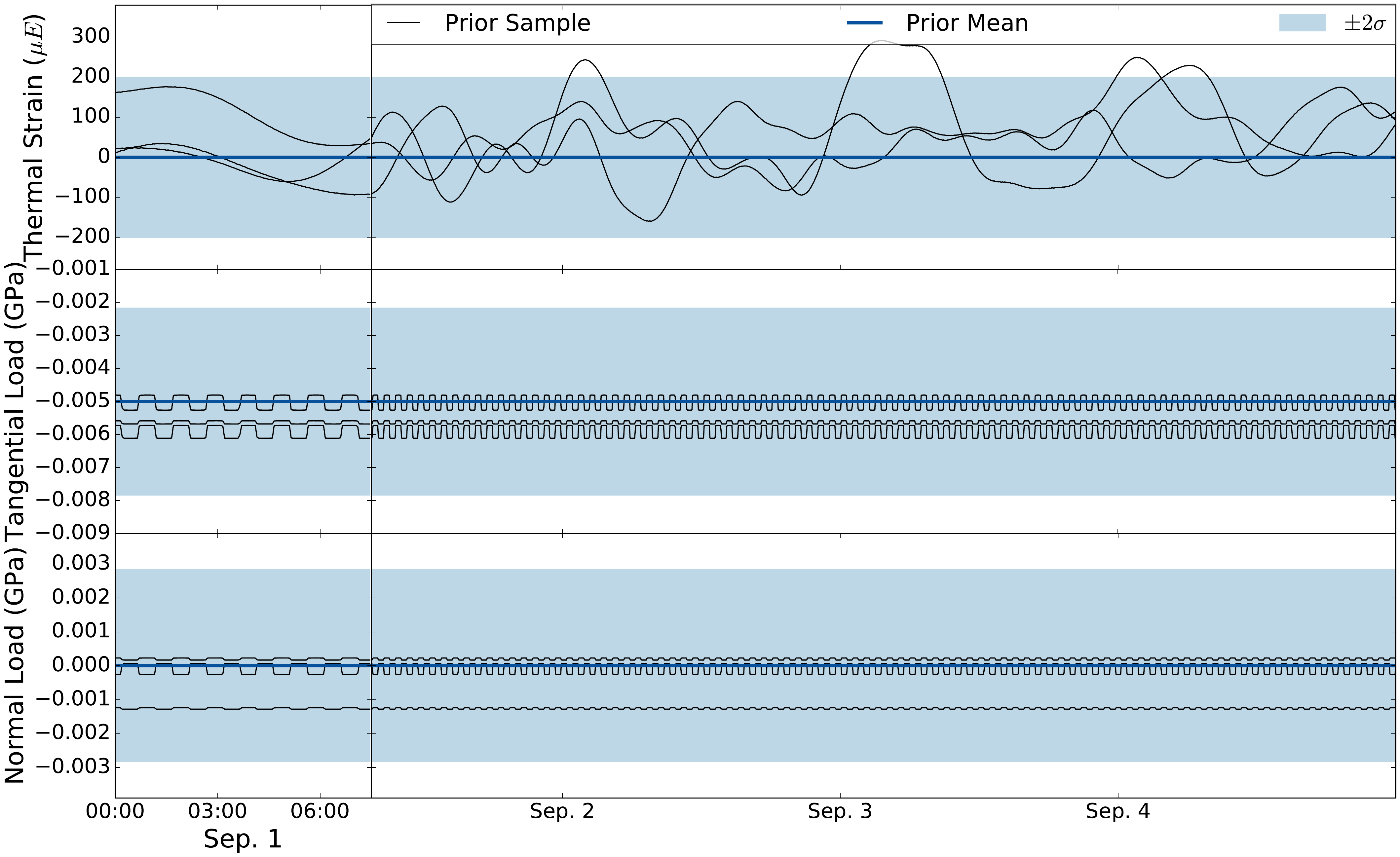}
\caption{Prior distribution and samples for beam example.  Even without incorporating any observational data, the prior samples have the general structure we would expect: the normal and tangential loads vary quickly because the water level (not plotted here) changes quickly, while the thermal strain varies more slowly over time.}
\label{fig:toyPrior}
\end{figure}

To complete the problem definition, we need to define the prior distributions on the normal $w_n(h^+)$ and tangential $w_t(h^+)$ loads.   Each of these loads is endowed with a Gaussian process prior with constant mean function and Matern covariance kernel.  For both $w_n$ and $w_t$, the covariance kernel has variance $\sigma^2=2\times 10^{12}$, lengthscale $L=5$, and smoothness parameter $\nu=5/2$.   The mean of the tangential force $w_t$ was chosen as $\num{-5e6}$ (Pa) and the mean of the normal force was chosen as $0.0$.  The negative tangential load indicates that we expect the tangential load to act against the hydrostatic load.   The prior on the loads $w_t$ and $w_n$ is shown in Figure \ref{fig:toyPrior}.

\begin{figure}[h!]
\centering
\includegraphics[width=\textwidth]{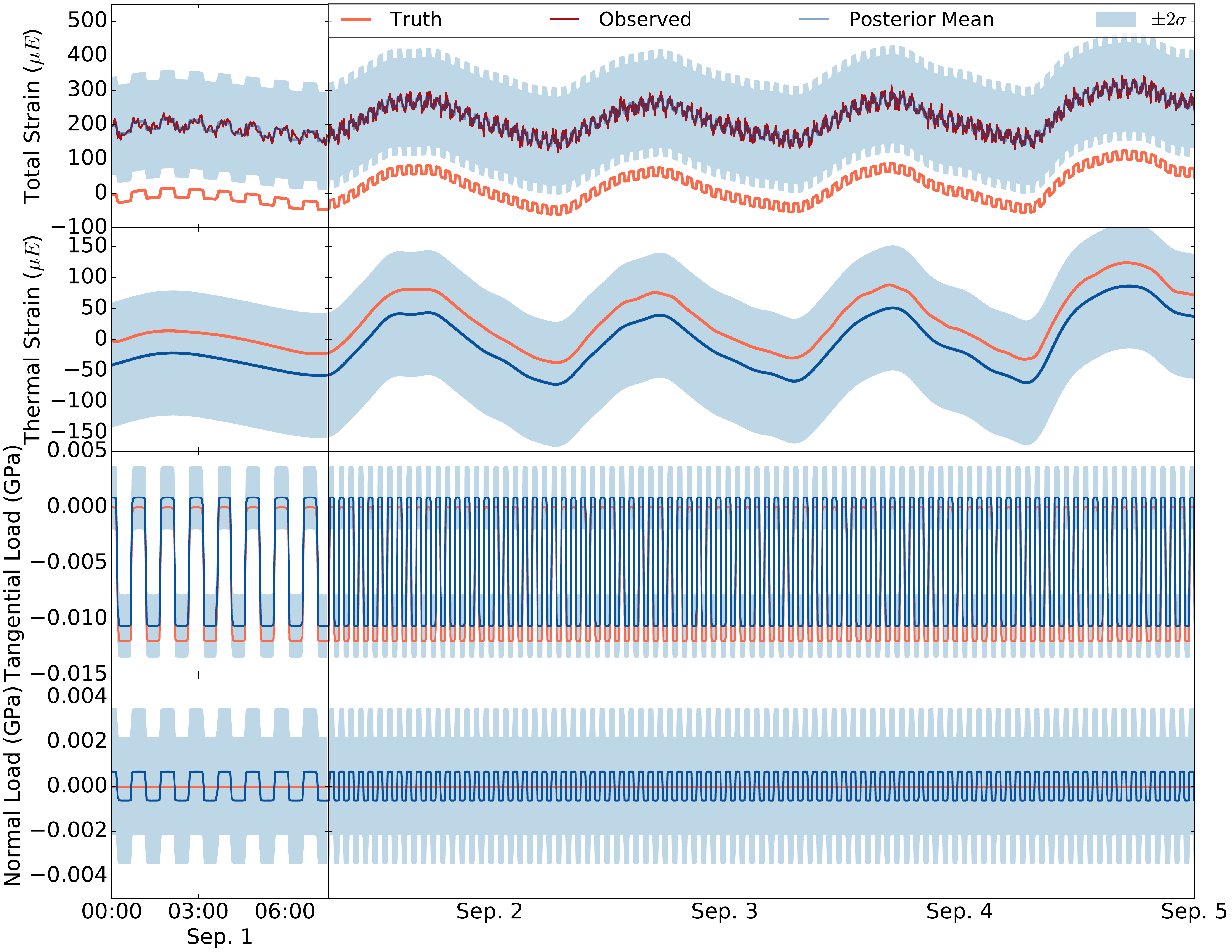}
\caption{Posterior distribution on simple beam problem.   The top row shows the posterior prediction on total strain as well as the true model field and observations.  The observations are generated by shifting and adding noise to the truth.  The next three rows show the posterior thermal strain, tangential loads, and normal loads. Overall, the posterior means match the truth quite well and the truth is always within the posterior credible intervals.}
\label{fig:toyPost}
\end{figure}

This illustrative problem is small enough that we can compute the posterior directly.  First, we evaluate the prior mean functions and covariance kernels at the times and water heights of interest.  This then gives us prior mean vectors and covariance matrices that we can use to analytically compute the posterior mean vector and covariance matrix.    The result is shown in Figure \ref{fig:toyPost}.   Note that our analysis characterizes the \textit{joint} distribution of $\bm{\varepsilon}_b$, $\bm{\varepsilon}_T$, $w_n$ and $w_t$, but only the marginal distributions are shown in Figure \ref{fig:toyPost}.

%
%

From the posterior results, it is clear that there are no daily cycles in the normal and tangential loads. This is because our forward model uses only the elastic portion of the strain which varies primarily with water height and applied hydrostatic load. Interestingly, the lack of thermal cycles in the posterior loads indicates that we can accurately account for thermal influences without direct temperature measurements. 

The posterior mean on the traction load is quite close to the truth and the posterior standard deviation encompasses the truth.   While the posterior mean for the normal load is not as close to the truth, the larger standard deviation indicates this inaccuracy that the truth is well within two standard deviations.  

Table \ref{tab:postOff} shows the posterior estimates of the strain gage offsets.  Despite the influences of thermal noise, gage biases, and random noise, our approach is still able to obtain an accurate characterization of the offsets.
\begin{table}[h!]
\centering
\caption{Posterior estimates and true strain-gage offsets.  The posterior mean is in good agreement with the truth, which is well within one standard deviation of the mean for all gages.}
\begin{tabular}{|c|rrr|}\hline
Gage Position & $x=(2,0.5)$ & $x=(6,0.5)$ & $x=(10,0.5)$\\\hline
True offset & $\num{2.0e-4}$ &$\num{-2.0e-4}$ & $\num{1.0e-4}$ \\
Posterior Mean & $\num{2.3e-4}$ & $\num{-1.7e-4}$ & $\num{1.2e-4}$\\
Posterior Std. Dev. & $\num{4.4e-5}$ & $\num{4.5e-5}$ & $\num{4.6e-5}$ \\\hline
\end{tabular}
\label{tab:postOff}
\end{table}

The results for this toy problem in Figure \ref{fig:toyPost} and Table \ref{tab:postOff} are impressive when considering the noise, calibration offsets, and thermal influences in the observations.   This indicates the potential power of our approach on the much larger lock gate problem.   However, the lock gate problem has many more parameters to infer and this increased dimensionality precludes the direct computation of the posterior mean and covariance.  Section \ref{sec:solution} will introduce a more sophisticated solution approach that scales linearly with the number of observations and can tackle the high dimensional lock gate problem.

\section{Inverse Problem Solution}\label{sec:solution}
As the number of observation times $N_{\text{obs}}$ and the number of boundary degrees of freedom $N_w$ grow, the memory requirements of directly computing the Gaussian posterior mean and covariance grow like $\mathcal{O}(N_\text{obs}^2 N_{w}^2)$ and the computational cost grows like $\mathcal{O}(N_\text{obs}^3 N_w^3)$.  Unfortunately, this is intractable for the the real lock gate problem, where we are getting tens of thousands of observations per week ($N_\text{obs} > 10^4$ ) and there are hundreds of parameters at each step ($N_w = 208$).  This means that for a single week of observations, approximately 465 GB of memory would be needed to simply store the posterior covariance matrix and solving the system would require on the order of $10^{18}$ floating point operations.  Clearly, a more sophisticated numerical approach is necessary to overcome this issue.  

Several approaches have recently been proposed for largescale linear-Gaussian problems, including optimal low rank approximations \cite{Spantini2015}, approximations with hierarchical matrices \cite{Borm2007,Litvinenko2017}, and representations of Gaussian processes through stochastic differential equations (SDE), both temporal \cite{Hartikainen2010} and spatial \cite{Lindgren2011}.  We will take the temporal SDE approach, because it allows us to use standard Kalman filtering approaches to process the data sequentially as it becomes available.

\subsection{Statespace formulation of Gaussian processes}\label{sec:statespace}
The Wiener-Khinchin theorem provides an interesting relationship between the spectral density of a random process and its covariance function.  Let $z(t)$ denote a stationary random process in $\mathbb{R}^{N_z}$ with covariance kernel $k_z(\tau): \mathbb{R} \rightarrow \mathbb{R}^{N_z\times N_z}$ defined by
\begin{equation}
k_z(\tau) = \mathbb{C}\text{ov}\left[z(t), z(t+\tau)\right],
\end{equation}
and mean function $\mu_z(t)$.  Let $S_z(\omega): \mathbb{R} \rightarrow \mathbb{R}^{N_z\times N_z}$ denote the matrix-valued spectral density of $z(t)$.  The Wiener-Khinchin theorem states that the covariance function is the inverse Fourier transform of the spectral density, so that
\begin{equation}
k_z(\tau) = \mathcal{F}^{-1}\left[S_z(\omega)\right]. \label{eq:wienerkhinchin}
\end{equation}

This relationship has been exploited in \cite{Hartikainen2010} and \cite{Solin2014} to construct linear time invariant (LTI) SDEs whose solutions are Gaussian processes with spectral densities corresponding to well known covariance kernels, like the Matern and quasi-periodic families of kernels.\footnote{We should also point out that \eqref{eq:wienerkhinchin} can be used to create an empirical covariance kernel from one or more realizations of a stationary random process \cite{Marcotte1996}.}   For a stationary covariance kernel $k_z(\tau)$ that admits such an SDE representation, it is possible to introduce a new state variable $v$ and matrices $F$, $L$, $H$ and $Q$ such that
\begin{eqnarray}
\frac{d v}{dt} &=& Fv(t) + Ly(t) \label{eq:sdesys}\\
z(t) &=& \mu_z(t) + H v(t), \label{eq:sdeobs}
\end{eqnarray}
where $v$ is a vector-valued state variable with dimension $N_v\geq N_z$,  $y(t)\sim N(0,Q)$ is white noise with spectral density $Q\in \mathbb{R}^{N_y\times N_y}$, $F\in \mathbb{R}^{N_v\times N_v}$ is system drift matrix, $L\in \mathbb{R}^{N_v\times N_y}$ is a matrix defining how the noise affects state evolution, and $H\in \mathbb{R}^{N_z\times N_v}$ is the observation matrix.  The initial condition is given by $v(0)\sim N(m_0, P_0)$.  Notice that $F$, $L$, $Q$, and $H$, as well as the initial covariance $P_0$ and the dimension of $v$, $N_v$, all depend on the form of $k_z(\tau)$.    The mean $m(t)$ and covariance $P(t)$ of the state variable $v$ satisfy the differential equations
\begin{eqnarray}
\frac{d m(t) }{dt} &=& F m(t)\\
\frac{d P(t) }{dt} &=& FP(t) + P(t)F^T + LQL^T, \label{eq:sdeform}
\end{eqnarray}
and the initial covariance satisfies the Lyapunov equation
\begin{equation}
FP_0 + P_0F^T + LQL^T = 0.
\end{equation}

While \cite{Hartikainen2010} and \cite{Solin2014} derived LTI SDEs for several families of covariance kernels, the more complicated covariance kernels in Sections \ref{sec:prior} and \ref{sec:error} contain combinations of the canonical forms studied in \cite{Hartikainen2010} and \cite{Solin2014}.  Fortunately, it is relatively straightforward to derive SDE representations for these concatenations and linear combinations of the canonical kernels.

To see this, let $z_i(t) \sim GP(\mu_i(t), k_i(\tau))$ be a Gaussian process with mean $\mu_i(t)$ and a stationary covariance kernel $k_i(\tau)$ for $i\in\{1,2,\ldots,M\}$.  Assume that each of these Gaussian processes admits an SDE representation with matrices $F^{(i)},L^{(i)},Q^{(i)},H^{(i)},$ and $P^{(i)}_0$.   Then the Gaussian process $z(t) =\sum_{i=1}^M \alpha_i z_i(t)$ admits an SDE representation with block diagonal matrices $F,L,Q,P_0$ and a block row matrix $H$.   The matrices $F,L,Q,P_0$ contain $F^{(i)},L^{(i)},Q^{(i)},$ and $P^{(i)}_0$ as blocks along their diagonals and $H$ is given by
\begin{equation}
H = \left[ \begin{array}{cccc} \alpha_1 H^{(1)} & \alpha_2 H^{(2)} & \ldots & \alpha_N H^{(N)} \end{array}\right].
\end{equation}
Using this expression, we can construct SDE representations for any linear combination of the canonical kernels studied in \cite{Hartikainen2010} and \cite{Solin2014} including coregional kernels.   This includes the thermal strain GP from \eqref{eq:thermal_gp_kernel} as well as the bias in \eqref{eq:gage_bias_gp}.   However, we cannot yet apply these SDE techniques to the boundary load Gaussian processes, which depend on water level and cannot be directly cast as a temporal SDE of the form in \eqref{eq:sdeform}.  We will employ Karhunen-Lo\'eve decompositions to circumvent this.

\subsection{Karhunen-Lo\'eve Decompositions}
Karhunen-Lo\'eve (KL) decompositions are a common parameter-reduction technique in Bayesian statistics (see e.g., \cite{Marzouk2009}).   The KL decomposition of a covariance kernel is an infinite dimensional analog of the singular value decomposition of a matrix.  In particular, the KL decomposition of a Gaussian process $w(x, \omega) \sim GP( \mu_w(x), k_w(x,x^\prime))$ defined for $x\in\Omega$ takes the form
\begin{equation}
w(x,\omega) = \mu_w(x) + \sum_{k=1}^\infty \sqrt{\lambda_k}\phi_k(x) z_k(\omega), \label{eq:generalKL}
\end{equation}
where each $z_k(\omega)$ is an independent standard normal random variable, and $\lambda_k$, $\phi_k$ satisfy the eigenvalue problem
\begin{equation}
\int_\Omega k_w(x,x^\prime) \phi_k(x^\prime) dx^\prime = \lambda_k \phi_k(x). \label{eq:klintegral}
\end{equation}
This integral equation can be solved efficiently using the the Nystrom method, which discretizes the integral and then solves a standard matrix eigenvalue problem.   The Nystrom method also provides a natural way of interpolating the discrete solution to evaluate $\phi_k(x)$ for any $x$.  More details of the Nystrom method can be found in Appendix \ref{sec:nystrom}.

The expansion in \eqref{eq:generalKL} has another important feature: it separates stochastic dependencies (i.e., dependence on the abstract random variable $\omega$) and deterministic dependencies (e.g., dependence on water levels $h^+$,$h^-$, or position $x$).   For tensor product covariance kernels, the KL decomposition also separates deterministic dependencies.   We will use this feature to transform the boundary load Gaussian process, which depends on $x,h^+,h^-$ and $t$, into a temporal SDE like those in the previous section.  More specifically, our goal is to represents the boundary load functions $w_q(x, h, t, \omega)$ and $w_m(x, h, t, \omega)$ as a linear combination of Gaussian processes that depend only on time.

Recall the tensor product form of the quoin covariance kernel in \eqref{eq:quoin_tensor}.  Taking advantage of this form, we can write the KL expansion of the quoin load Gaussian process $\quoin(x, h^{+}, h^{-}, t, \omega)$ as
\begin{equation}
\quoin(x, h, t, \omega) =  \mu_q(x, h,t) + \sum_{i=1}^\infty \alpha_{xi}\sqrt{\lambda_{xi}} \phi_{xi}(x)\left[\sum_{j=1}^\infty \alpha_{hj}\sqrt{\lambda_{hj}}\phi_{hj}(h)z_{j}(t,\omega)\right], \label{eq:kl1}
\end{equation}
where $\lambda_{xi}$ and $\phi_{xi}(x)$ are eigenvalues and eigenfunctions for $k_{qx}$, $\lambda_{hj}$ and $\phi_{hj}(h)$ are eigenvalues and eigenfunctions for $k_{qh}$, and $z_{hj}(t) \sim GP(0, k_{qt}(t_1,t_2))$ is a vector-valued Gaussian process with identity marginal covariance.

After truncating the sums in \eqref{eq:kl1}, we are left with 
\begin{eqnarray}
\quoin(x, h, t, \omega) & =&  \mu_q(x, h,t) + V_{qx}(x)V_{qh}(h) z_{q}(t, \omega), \label{eq:contquoin}
\end{eqnarray}
where $V_{qx}(x)$ is a row vector with components $\alpha_{xi}\sqrt{\lambda_{xi}} \phi_{xi}(x)$ and $V_{qh}(h)$ is a row vector with components $\alpha_{hj}\sqrt{\lambda_{hj}}\phi_{hj}(h)$.  A similar expansion and truncation can be performed using the miter covariance kernel in \eqref{eq:miter_tensor} to obtain
\begin{eqnarray}
\miter(x, h, t, \omega) & =&  \mu_m(x, h,t) + V_{mx}(x)V_{mh}(h) z_{m}(t, \omega). \label{eq:contmiter}
\end{eqnarray}
Notice that the stochastic dependence on $\omega$ is now limited to $z_{q}(t, \omega)$ and $z_{m}(t, \omega)$, which are Gaussian processes that only depend on time.

The expressions in \eqref{eq:contquoin} and \eqref{eq:contmiter} describe the continuous boundary loads at any spatial location $x$.  These can be projected on to the finite element basis functions to obtain
\begin{eqnarray}
\quoinvec(h,t,\omega) & = & \bm{\mu}_q(h,t) + \bm{V}_{qx}V_{qh}(h) z_{q}(t, \omega) \label{eq:quoinexpand}\\
\mitervec(h,t,\omega) & = & \bm{\mu}_m(h,t) + \bm{V}_{mx}V_{mh}(h) z_{m}(t, \omega), \label{eq:miterexpand}
\end{eqnarray}
where $\bm{V}_{qx}$ is a matrix with rows containing the projection of $V_{qx}(x)$ on to each finite element basis function.  If the time kernels $k_{qt}$ and $k_{mt}$ allow for SDE representations of $z_{q}(t,\omega)$ and $z_m(t,\omega)$, then  \eqref{eq:quoinexpand} and \eqref{eq:miterexpand} can be used to represent the load vectors $\quoinvec(h, t, \omega)$ and $\mitervec(h, t, \omega)$ as SDEs as well, which allows us to employ efficient Kalman filtering and smoothing techniques to characterize the posterior.  Like \cite{Hartikainen2010}, we use sparse linear algebra routines, a standard Kalman filter, and a Rauch-Tung-Striebel smoother.  Appendix \ref{sec:app:kalman} provides more details for the Kalman smoother formulation.

\section{Miter gate results}\label{sec:results}
Here we test our formulation and statespace solution approach on a real USACE miter gate on Mississippi River Lock 27 near St. Louis.   A detailed finite element model with approximately $3\times 10^5$ degrees of freedom was reduced using the static condensation approach in Section \ref{sec:model:cond} to obtain a system with $1175$ degrees of freedom pertaining to the boundaries and regions near $14 $ different strain gages.    The quoin boundary contains $281$ degrees of freedom and the miter boundary contains $280$, for a total of $461$ model parameters at every point in time.   Observations from the SMART Gate database were obtained at $2200$ different times between April 14th and April 16th, 2014.  Combined with $14$ parameters describing the thermal strain at every time and $14$ static bias terms, this implies that the problem has 1,045,000 parameters.  Even after a significant parameter reduction from the KL decomposition (35 modes for both the quoin and miter), there are still 184,800 parameters, which is significantly larger than what we can tackle directly.

The prior on the loads is adopted from Section \ref{sec:prior}.  In particular, the kernels in \eqref{eq:quoin_tensor} and \eqref{eq:miter_tensor} are constructed from a Matern kernel with $\nu=3/2$, $L = 10$ inches, and $\sigma^2=2$, and $\beta=5$.  These values were chosen based on the expected range of boundary forces as well and engineering judgement for a feasible lengthscale.  Future work investigating these settings would be useful before deploying this SHM system in practice.  For both the quoin and miter, 35 KL modes were used to capture 95\% of the energy from $k_{qx}$ and $k_{mx}$.    The water height kernels $k_{qh}$ and $k_{mh}$ were both chosen to be  Matern kernels with $\nu=5/2$, $L=120$ inches, and $\sigma^2=1$.  This long lengthscale reflects our belief the loads vary almost linearly with water level.   The thermal kernels are of the form in \eqref{eq:thermal_gp_kernel} and have the parameters in Table \ref{tab:thermalParams}.   Moreover, the variance of the observation noise $\epsilon(t)$ was set to $9$, which is representative of the noise observed in the strain gages under constant loading conditions.

\begin{table}[h!]
\caption{Thermal covariance kernels used in Lock 27 miter gate problems.  All kernels are of the quasi-periodic form in \eqref{eq:thermal_gp_kernel}.  Gages near the bottom of the gate are given a lower variance than gages near the top of the gate because the bottom gages are either fully or partially submerged much of the time and are therefore less susceptible to temperature fluctuations.  All periods and lengths are in units of days.}
\begin{tabular}{|ccc|ccc|ccc|ccc|}\hline
Gage & Girder & Location & \multicolumn{3} {c|} {$\kappa_{T1}$} & \multicolumn{3} {c|} {$\kappa_{T2}$} & \multicolumn{3} {c|} {$\kappa_{T3}$} \\\hline
 &&& $P$ & $L$ & $\sigma^2$ & $\nu$ & $L$ & $\sigma^2$ & $\nu$ & $L$ & $\sigma^2$\\\hline
0   & 3  & 3    & 1 & 0.5 & 1.0 & 1.5 & 0.5 & 225 & 1.5 & 28 &  900 \\
1   & 3  & 11  & 1 & 0.5 & 1.0 & 1.5 & 0.5 & 225 & 1.5 & 28 & 900\\\hline
2   & 5  & 6    & 1 & 0.5 & 1.0 & 1.5 & 0.5 & 225 & 1.5 & 28 & 900\\
3   & 5  & 19  & 1 & 0.5 & 1.0 & 1.5 & 0.5 & 225 & 1.5 & 28 & 900\\\hline
4   & 7  & 6    & 1 & 0.5 & 1.0 & 1.5 & 0.5 & 225 & 1.5 & 28 & 900\\
5   & 7  & 19  & 1 & 0.5 & 1.0 & 1.5 & 0.5 & 25 & 1.5 & 28 & 400\\\hline
6   & 9  &   6  & 1 & 0.5 & 1.0 & 1.5 & 0.5 & 25 & 1.5 & 28 & 400\\
7   & 9  & 19  & 1 & 0.5 & 1.0 & 1.5 & 0.5 & 25 & 1.5 & 28 & 400\\\hline
8   & 11 &  9  & 1 & 0.5 & 1.0 & 1.5 & 0.5 & 25 & 1.5 & 28 & 400\\
9   & 11 & 26 & 1 & 0.5 & 1.0 & 1.5 & 0.5 & 25 & 1.5 & 28 & 400\\\hline
10 & 12 &   9 & 1 & 0.5 & 1.0 & 1.5 & 0.5 & 2.5 & 1.5 & 28 & 100\\
11 & 12 & 26 & 1 & 0.5 & 1.0 & 1.5 & 0.5 & 2.5 & 1.5 & 28 & 100\\\hline
12 & 13 & 12 & 1 & 0.5 & 1.0 & 1.5 & 0.5 & 2.5 & 1.5 & 28 &100\\
13 & 13 & 32 & 1 & 0.5 & 1.0 & 1.5 & 0.5 & 2.5 & 1.5 & 28 & 100\\\hline
\end{tabular}
\label{tab:thermalParams}
\end{table}

\begin{figure}[h]
\centering
\includegraphics[width=\textwidth]{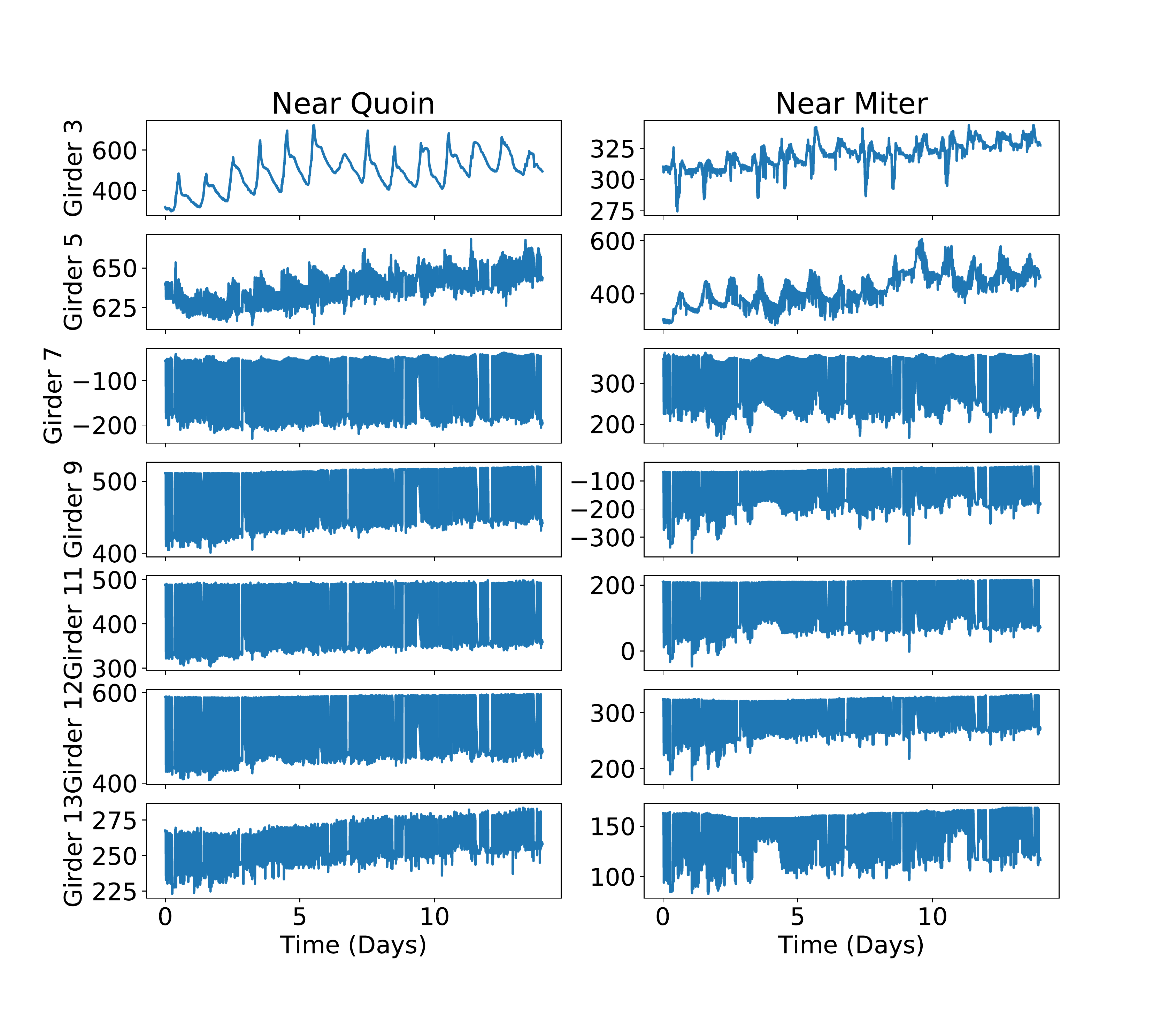}
\caption{Observed strain $\hat{\bm{\strain}}_{\text{obs}}(t)$ for Lock 27 over approximately two weeks in April 2014.  Each plot contains a time series of observations for a single strain gage.  The gage locations can be found in Figure \ref{fig:miterLoadPosterior}.   In general, the diurnal cycles are less obvious for gages that are lower on the gate, which are partially or fully submerged and less susceptible to temperature fluctuations in the air.}
\label{fig:rawStrains}
\end{figure}

\begin{figure}[h]
\centering
\begin{subfigure}[t]{\textwidth}
\includegraphics[width=\textwidth]{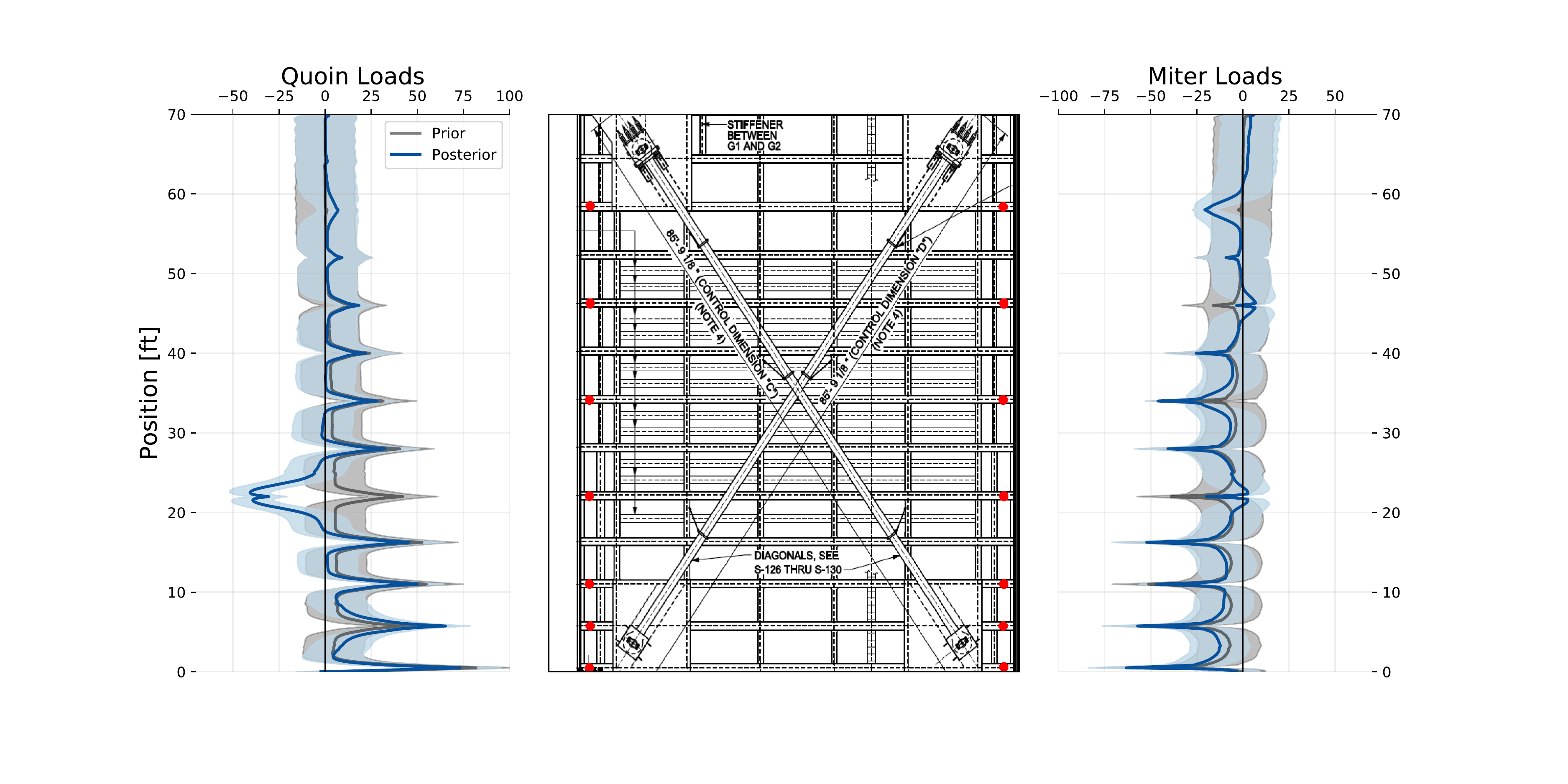}
\caption{Posterior loads when the water levels on either side of the gate are substantially different.  In this regime, there is enough water pressure to effectively seal the miter-miter interface and our structural model is a decent representation of the reality and it is possible to draw conclusions from the posterior.  The posterior is similar in shape to prior except on the quoin near girder 9, where an unphysical negative load occurs.  This could be a result of a modeling inaccuracy near that point, a faulty strain gage on that girder, or a potential problem with the lock gate itself.  Further investigation is needed to identify the true cause of this difference.}
\end{subfigure}

\begin{subfigure}[t]{\textwidth}
\includegraphics[width=\textwidth]{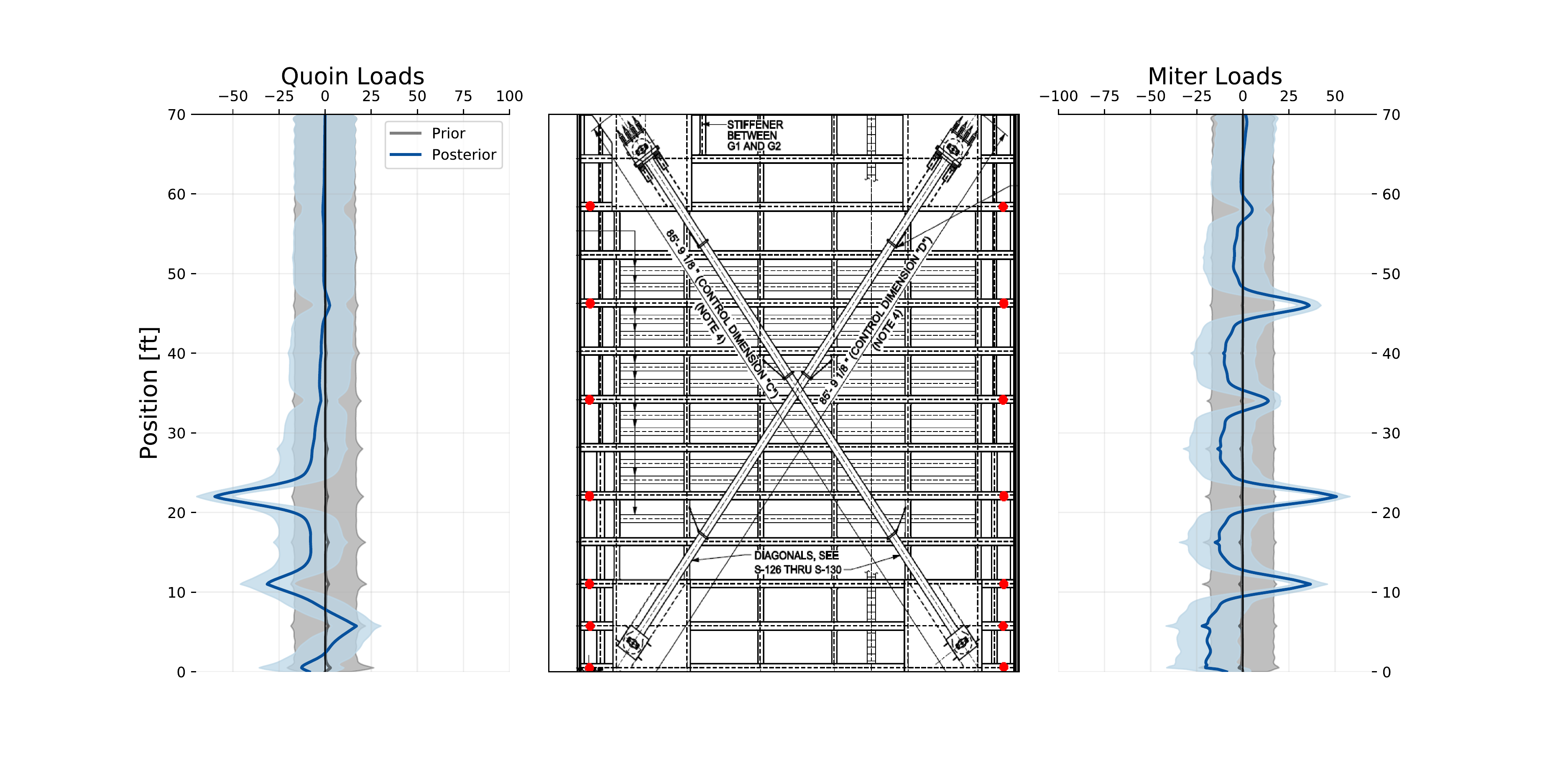}
\caption{Posterior loads when the water levels are the same.  In this situation, the miter-miter interface is not well sealed and our physical model is inaccurate.  This is clear in the posterior, where the significant difference from the prior does not have a physically meaningful explanation.} 
\end{subfigure}

\caption{Snapshot of prior and posterior boundary loads at a single time as well as a plan view of the the miter gate used in this study.  Red dots on the diagram show locations of the $14$ strain gages.  As expected, the loads are larger in magnitude at girder locations.   The posterior variance is also larger for girders that do not have strain gages, such as girders 8 and 10 (counted from the top).}
\label{fig:miterLoadPosterior}
\end{figure}

\begin{figure}[h]
\centering
\begin{subfigure}[t]{\textwidth}
\centering
\includegraphics[width=0.8\textwidth]{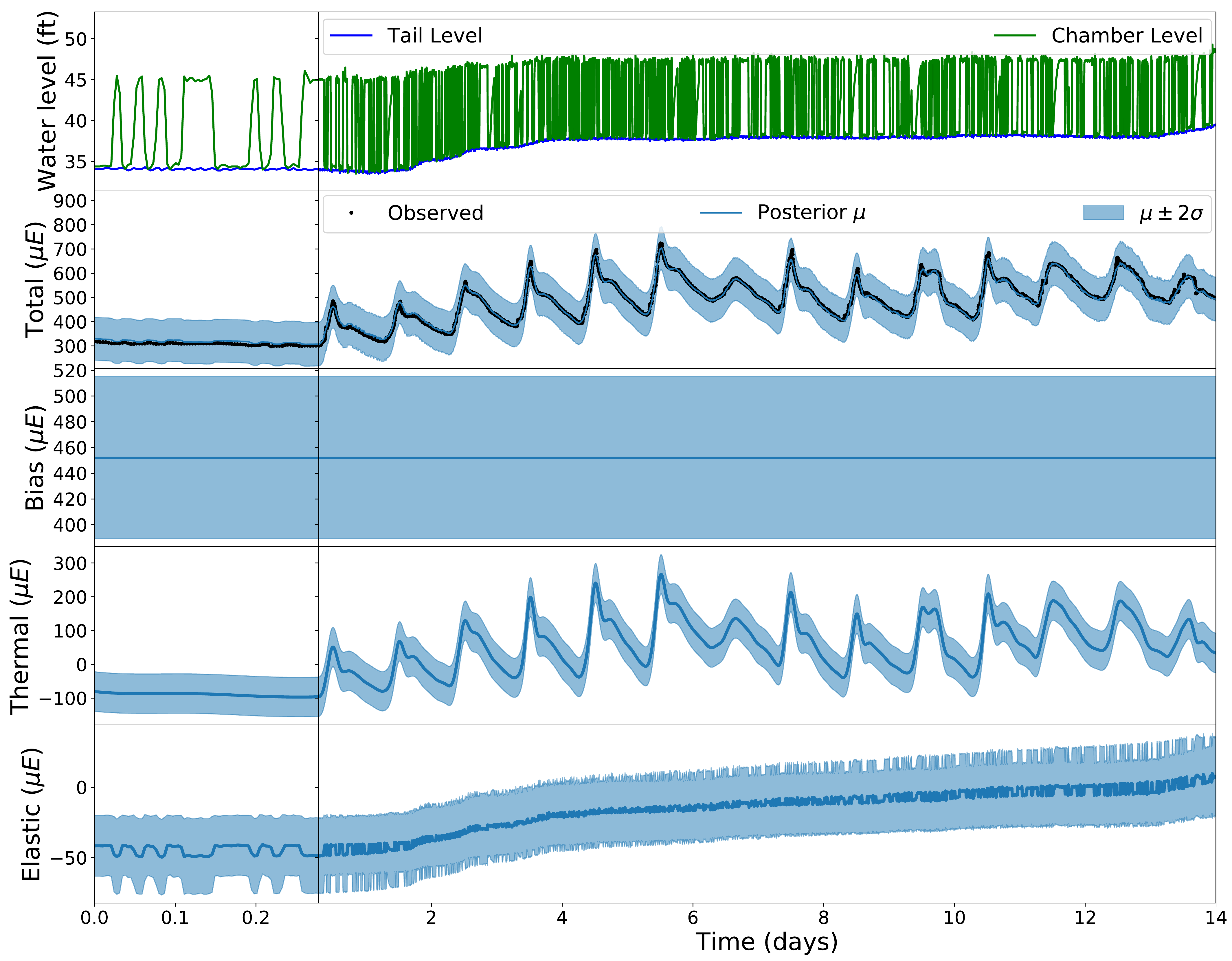}
\caption{Strains at gage 0 located at the top of the gate near the Quoin.  This gage is dominated by thermal strains, but our approach is still able to determine the small elastic strains that result when the lock chamber is filled and drained.}
\end{subfigure}
\begin{subfigure}[t]{\textwidth}
\centering
\includegraphics[width=0.8\textwidth]{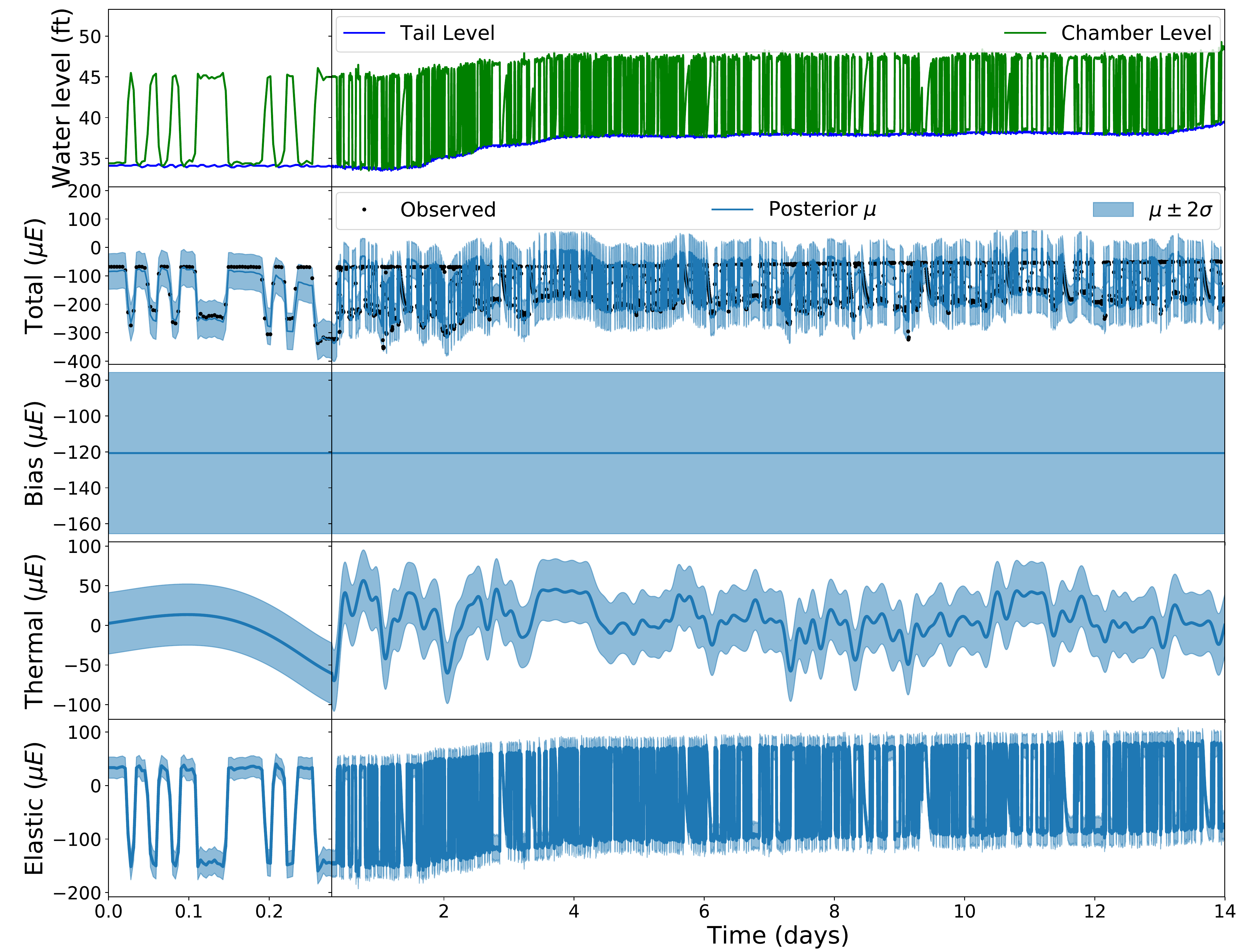}
\caption{Strains at gage 7 located in the middle of the gate near the Miter.}
\end{subfigure}

\caption{Posterior predictive distributions over strain components for two gages on Lock 27.  The chamber level corresponds to the upstream water level $h^+$ and the tail level to downstream water level $h^-$.}
\label{fig:miterStrainPosterior}
\end{figure}

Figure \ref{fig:rawStrains} shows observed strains from the SMART Gate database taken every minute for approximately a week in April 2014.  These data will be used to learn the boundary loads, thermal strains, and gage biases.   Notice that there are distinct diurnal cycles in several of the gages.  These gages are typically near the top of the gate and experience the largest air temperature and solar irradiance fluctuations.   Figure \ref{fig:miterLoadPosterior} shows the posterior loads computed with this data at select times and Figure \ref{fig:miterStrainPosterior} shows the posterior over the thermal strain, gage bias, and posterior predictive elastic strain for a select number of gages.   Note that over this two week interval, the posterior loads do not depend directly on time, they only depend on the upstream and downstream water levels $h^+$ and $h^-$.   Thus, the plots in figure \ref{fig:miterLoadPosterior} are representative of the posterior at many different times.   

The computed posterior exists over all loads, biases, and thermal strains for all times, but only a few marginal snapshots are shown here due to limited space.  Future analyses could use information from the entire joint posterior distribution and an interactive website has been developed using Bokeh \cite{bokeh} that allows users to explore the full posterior.

\section{Conclusions}\label{sec:conclusions}
We have introduced a novel combination of techniques, including static condensation, statespace representations of Gaussian processes, Karhunen-Loeve decompositions, and Kalman smoothing, that enables large structural health monitoring problems to be solved efficiently.   
The use of Gaussian processes makes our approach flexible and capable of capturing complex environmental influences.  However, the efficiency and scalability of our approach relies on the statespace representation of all Gaussian process priors.  Transforming the Gaussian processes into time-dependent stochastic differential equations enabled us to perform inference efficiently using Kalman filtering and smoothing techniques.  Using these techniques, our methodology scales linearly with the number of observations.

Our approach was able to combine a high fidelity finite element simulation of a gate on Lock 27 near St. Louis Missouri with real strain observations stored in the SmartGate database.  The observations contain significant environmental influences that manifested as a cyclic thermal strain.  We have shown that quasi-periodic Gaussian processes are capable of capturing these unmodeled environmental effects, allowing us to characterize a model that does not explicitly account for thermal expansion.  Thermal influences are ubiquitous in strain gages and we believe that similar techniques will be useful in a broad range of structural health monitoring and data analysis problems.

While our examples here used a fixed amount of data, this is not required.  In fact, using just a Kalman filter, and not a Kalman smoother, would allow our approach to easily processing streaming datasets in near real-time.  For example, strain gage observations could be incorporated as they become available to get an accurate picture of the latest conditions on the gate.  This would enable lock gate operators to frequently assess the gate's health and rapidly respond to potential problems.

\section*{Acknowledgments}
The authors would like to thank Travis Fillmore, Steven Bunkley, and Quincy Alexander for their work setting up the structural model and computing the reduced system matrices described in Section \ref{sec:model:cond}.  We would also like to thank Brian Eick for helping extract the strain gage data in Section \ref{sec:results} and Arnold Song for helping construct figures in Section \ref{sec:results}.

\bibliographystyle{siamplain}
\bibliography{references}

\appendix
\section{Nystrom Method}\label{sec:nystrom}
To compute the KL decompositions and provide a method for evaluating $V_{mx}(x)$ and $V_{mh}(h)$ at any input location, we employ the Nystrom method.  Consider a general random process $u(y)$ with input (e.g., water level) $y\in\Omega_y\subseteq \mathbb{R}^{N_y}$ and covariance kernel $\kappa_u(y,y^\prime)$ and let $\phi_{uk}(y)$ denote the $k^{th}$ modes of the Karhunen-Lo\'eve decomposition.   For some eigenvalue $\lambda_{uk}$, $\phi_{uk}(y)$ satisfies 
\begin{equation}
\int_{\Omega_y} \kappa_u(y,y^\prime) \phi_{uk}(y) dy = \lambda_{uk} \phi_{uk}(y^\prime), \label{eq:eigenProblem}
\end{equation}
where $\lambda_{uk}$ and $\phi_{uk}(y)$ are the $k^{\text{th}}$ eigenvalue and eigenfunction of the covariance kernel $\kappa_u(y,y^\prime)$.  Approximating the integral with a quadrature rule\footnote{In our examples, we simple left-point rules for integration, resulting in equal weights across all quadrature points.} containing $M<\infty$ seed points $\{y^{(1)}, y^{(2)}, \ldots, y^{(M)}\}$ and weights $\{w^{(1)}, w^{(2)}, \ldots, w^{(M)}\}$ leads to the expression
\begin{equation}
\sum_{m=1}^M w^{(m)}\kappa_u(y^{(m)},y^\prime) \phi_{uk}(y^{(m)}) \approx \lambda_{uk} \phi_{uk}(y^\prime).\label{eq:nystromapprox}
\end{equation}
Using the same points to discretize $y^\prime$ yields the discrete eigenvalue problem
\begin{eqnarray}
\sum_{m=1}^M w^{(m)}\kappa_u(y^{(m)},y^{(i)}) \phi_{uk}(y^{(m)}) &=& \lambda_{uk} \phi_{uk}(y^{(i)})\\
\Rightarrow W K \bm{\phi}_{uk} &=& \lambda_{uk} \bm{\phi}_{uk}, \label{eq:nystromgeneig}
\end{eqnarray}
where $W$ is a $M\times M$ diagonal matrix containing the weights $w^{(m)}$, $K$ is an $M\times M$ covariance matrix containing the covariance of $u$ at the seed points, and $\bm{\phi}_{uk}$ is the vector containing evaluations of the $k^{th}$ KL mode at the seed points.  This method of discretizing the eigenvalue problem in \eqref{eq:eigenProblem} is called the Nystrom method.

Numerically, it is more efficient and accurate to solve symmetric eigenvalue problems than the non-symmetric problem in \eqref{eq:nystromgeneig}.  To symmetrize \eqref{eq:nystromgeneig}, consider a new variable $v_k = W^{-1/2} \bm{\phi}_{uk}$.  Multiplying \eqref{eq:nystromgeneig} on the left by $W^{-1}$ and rewriting with $v_k$ yields
\begin{eqnarray}
\Gamma \bm{\phi}_{uk} &=& \lambda_{uk} W^{-1}\bm{\phi}_{uk},\\
\Gamma W^{1/2} v_k &=& \lambda_{uk} W^{-1/2}v_k,\\
W^{1/2} \Gamma W^{1/2} v_k & = & \lambda_{uk} v_k, \label{eq:nystromsymeig}
\end{eqnarray}
which is a symmetric positive definite eigenvalue problem that can be solved efficiently.  The original eigenvector $\bm{\phi}_{uk}$ can then be easily calculated using the identity $\bm{\phi}_{uk} = W^{1/2}v_k$.

In \eqref{eq:contquoin} and \eqref{eq:contmiter}, we need to evaluate the KL modes at any point, not just at the quadrature points used in \eqref{eq:nystromgeneig}.  Fortunately, the integral discretization in \eqref{eq:nystromapprox} also defines an interpolation function.  Rearranging, we have 
\begin{equation}
\phi_{uk}(y^\prime) \approx \frac{1}{\lambda_{uk}} \sum_{m=1}^M w^{(m)}\kappa_u(y^{(m)},y^\prime) \phi_{uk}(y^{(m)}).\label{eq:nystrominterp}
\end{equation}
Thus, to approximation the KL mode $\phi_{uk}(y^\prime)$ and any point $y^\prime$, we first solve the symmetric eigenvalue problem from \eqref{eq:nystromsymeig}, then compute the cross covariance with the seed points $\kappa_u(y^{(k)},y^\prime)$, and finally evaluate the sum in \eqref{eq:nystrominterp} to obtain $\phi_{uk}(y^\prime)$.

\section{Kalman Smoother Formulation}\label{sec:app:kalman}
Our goal is to write each component of strain, $\strainvec_T(t,\omega)$, $\strainvec_b(\omega)$, and $\strainvec_{er}(h,t,\omega)$, using the statespace approaches described in Section \ref{sec:statespace}.   Once the statespace form is constructed, we can interpret the finite element model from Section \ref{sec:elasticGP} as a linear observation operator, thus allowing us to formulate the inference problem as filtering problem and apply Kalman filtering and smoothing techniques that have linear complexity in the number of observation times.

First, consider the thermal strain $\strainvec_T(t,\omega) $ from \eqref{eq:thermalForm}.  Each of the independent $z_{T,i}(t, \omega)$ random variables is a quasi periodic Gaussian process that we can express through an SDE of the form
\begin{eqnarray}
\frac{d v_{T,i}}{dt} &=& F_{T,i} v_{T,i}(t,\omega) + L_{T,i} y_{T,i}(t,\omega) \label{eq:thermalSDE} \\ 
z_{T,i} &=& H_{T,i} v_{T,i}
\end{eqnarray}
Collecting $F_{T,i}$, $L_{T,i}$, and $H_{T,i}$ into block diagonal matrices $F_T$, $L_T$, and $H_T$ allows us to write 
\begin{eqnarray}
\frac{d v_T}{d_t} &=& F_T v_T(t,\omega) + L_T y_T(t,\omega)\\
z_T(t,\omega) &=& H_T v_T(t,\omega), \label{eq:thermalSDE_Obs}
\end{eqnarray}
where $v_T$, $y_T$, and $z_T$ are the concatenations of the components $v_{T,I}$, $y_{T,i}$, $z_{T,I}$.  Using this expression, the thermal strain $\strainvec_T(t,\omega) $ can then be expressed as 
\begin{equation}
\strainvec_T(t,\omega) = \mu_T(t) + \sqrt{\Sigma_T} \, H_T\, v_T(t,\omega)
\end{equation}
Thus, when $\strainvec_T(t,\omega)$ has $N$ components, we will need to solve $N$ SDEs of the form in \eqref{eq:thermalSDE} to simulate realizations the thermal strain.   Notice that to handle the quasi-periodic structure of $z_{T,i}$, the dimension of the state $v_{T,I}$ is generally much larger than the scalar $z_{T,i}$ and the dimension of $v_T(t,\omega)$ can become quite large.

The bias terms $\strainvec_b(\omega)$ are constant in time and can thus be written trivially using an SDE of the form
\begin{eqnarray}
\frac{dv_b}{dt} &=& 0\\
\strainvec_b(\omega) &=& \mu_b + H_b v_b(t,\omega), \label{eq:biasSDE_Obs}
\end{eqnarray}
where $H_b=I$ is the identity. Note that only the initial conditions $v_b(t=0,\omega)$ are stochastic with zero mean and covariance $\Sigma_b$.

The elastic strain $\strainvec_{er}(h,t,\omega)$ can also be represented in statespace form using a KL decomposition to help define a height-dependent $H$ matrix.   Recall the KL expansions from \eqref{eq:quoinexpand} and \eqref{eq:miterexpand} and the time dependent random variables $z_q(t,\omega)$ and $z_m(t,\omega)$.  Each component $z_{q,i}(t,\omega)$ and $z_{m,i}(t,\omega)$ is a scalar Gaussian process with mean zero and covariance kernel $k_{qt}$.  We have chosen the covariance kernel to be of Matern class, which enables us to express the components in statespace form as
\begin{eqnarray}
\frac{d v_{q,i}}{dt} &=& F_{q,i} v_{q,i}(t,\omega) + L_{q,i} y_{q,i}(t,\omega)\\
z_{q,i}(t,\omega) &=& H_{q,i} v_{q,i}(t,\omega),
\end{eqnarray}
for appropriately chosen matrices $F_{q,i}$, $L_{q,ki}$, and $H_{q,i}$.   Similar to the thermal strain above, block diagonal matrices $F_q$, $L_q$, and $H_q$ can be created to obtain a system of the form
\begin{eqnarray}
\frac{d v_{q}}{dt} &=& F_{q} v_{q}(t,\omega) + L_{q} y_{q}(t,\omega)\\
z_{q}(t,\omega) &=& H_{q} v_{q}(t,\omega).
\end{eqnarray}
Performing a similar procedure for the miter loads, we can express the elastic strain as
\begin{equation}
\strainvec_{er}(h,t,\omega) =  \strainvec_{er,w}(h) + \bm{B}_r\bm{K}_r^{-1} \left[ \begin{array}{c} \bm{\mu}_q(h,t) + \bm{V}_{qx}V_{qh}(h) H_q v_q(t,\omega)\\  \bm{\mu}_m(h,t) + \bm{V}_{mx}V_{mh}(h) H_m v_m(t,\omega) \end{array}\right] . \label{eq:elasticSDE_Obs}
\end{equation}
We can now combine $v_T$, $v_b$, $v_q$, and $v_m$ into a single state vector $v$ that takes the form
\begin{equation}
v(t,\omega) = \left[\begin{array}{c} v_{T}(t,\omega) \\ v_{b}(\omega)\\ v_{q}(t,\omega)\\ v_{m}(t,\omega)\end{array}\right]
\end{equation}
The white noise processes can also be combined to obtain
\begin{equation}
y(t,\omega) = \left[\begin{array}{c} y_{T}(t,\omega)\\ y_{q}(t,\omega)\\ y_{m}(t,\omega)\end{array}\right].
\end{equation}
Thus, we are left with an SDE of the form
\begin{equation}
\frac{d v}{dt} = F v(t,\omega) + L y(t,\omega), \label{eq:fullSDE}
\end{equation}
where,
\begin{equation}
F = \left[ \begin{array}{cccc} F_T & 0 & 0 & 0\\ 0 & 0 & 0 & 0 \\ 0 & 0 & F_q & 0 \\ 0 & 0 & 0 & F_m \end{array}\right]
\end{equation}
and
\begin{equation}
L = \left[ \begin{array}{ccc} L_T & 0 & 0 \\ 0 & 0 & 0 \\ 0 & L_q & 0 \\ 0 & 0 & L_m \end{array}\right].
\end{equation}

The observation operators in \eqref{eq:thermalSDE_Obs}, \eqref{eq:biasSDE_Obs}, and \eqref{eq:elasticSDE_Obs} can also be combined into a single expression of the form
\begin{equation}
\bm{\strain}_{\text{obs}}(t,\omega) = \mu_{\text{obs}}(t) + H v(t,\omega) + \epsilon, \label{eq:fullSDE_Obs},
\end{equation}
where $\mu_{\text{obs}}(t)$ and $H$ come from a concatenation of  \eqref{eq:thermalSDE_Obs}, \eqref{eq:biasSDE_Obs}, and \eqref{eq:elasticSDE_Obs}.  Our problem can now be cast as inferring $v(t,\omega)$ given observations of $\bm{\strain}_{\text{obs}}(t,\omega)$ and the model defined by \eqref{eq:fullSDE}--\eqref{eq:fullSDE_Obs}.   Once the posterior on $v(t,\omega)$ is characterized, we can reconstruct the posterior on the thermal strain $\strainvec_T(t,\omega)$, the bias terms $\strainvec_b(t,\omega)$, and the boundary loads $\bm{w}_q(t,\omega)$ and $\bm{w}_m(t,\omega)$.

Recall that the observational data at time $t_i$ is denoted by $\hat{\bm{\strain}}_{\text{obs}}(t_i)$ and is treated as a realization of the random variable $\bm{\strain}_{\text{obs}}(t_i,\omega)$.  Assume there are $M$ observation times and we are interested in the posterior over $v(t_i, \omega)$ given all observations time, i.e.,
\begin{equation}
 \pi\left( v(t_i,\omega) | \hat{\bm{\strain}}_{\text{obs}}(t_1), \hat{\bm{\strain}}_{\text{obs}}(t_2), \ldots, \hat{\bm{\strain}}_{\text{obs}}(t_M)\right). \label{eq:marginalPost}
 \end{equation}
 
 Notice that the SDE in \eqref{eq:fullSDE} defines a linear time invariant system for the state variable $v(t,\omega)$.  From this LIT system, a discrete Markov process can be defined so that 
 \begin{eqnarray}
 v(t_{i}, \omega) &=& \bar{F}_i v(t_{i-1}) + \bar{y}(t_{i-1}) \label{eq:markov1} \\
\bm{\strain}_{\text{obs}}(t,\omega) &=& \mu_{\text{obs}}(t) + H v(t,\omega) + \epsilon, \label{eq:markov2}
 \end{eqnarray}
 where $\bar{y} \sim N(0, \bar{Q}_{i-1})$ and 
 \begin{eqnarray}
 \bar{F}_i &=& \Phi(t_{i}-t_{i-1})\\
 \bar{Q}_{i-1} &=& \int_0^{t_{i}-t_{i-1}} \Phi( t_{i}-t_{i-1} - \tau) LQL^t  \Phi( t_{i}-t_{i-1} - \tau)^T d\tau, \label{eq:markovCov1}
 \end{eqnarray}
 where $\phi(x) = \exp(F x)$ is the matrix exponential of $Fx$, and the matrices $L$, $Q$, $H$ and $F$ are defined in \eqref{eq:fullSDE}.  See \cite{Hartikainen2010} for a deeper discussion of \eqref{eq:markov1}--\eqref{eq:markovCov1}.  With the discrete time Markov process in \eqref{eq:markov1}, the problem of characterizing \eqref{eq:marginalPost} for all times $t_i$ can be naturally cast as a combination of filtering and smoothing problems. 

\paragraph{Filtering}
The goal of filtering is to characterize the density $\pi( v(t_i) | \hat{\bm{\strain}}_{\text{obs}}(t_1), \ldots, \hat{\bm{\strain}}_{\text{obs}}(t_i) )$.
Let $\mu(t_i)$ and $\Sigma(t_i)$ denote the mean and covariance defining the density $\pi( v(t_i) |  \mu_{\text{obs}}(t_1), \ldots, \mu_{\text{obs}}(t_i) )$, i.e., the mean and covariance of $v(t_i)$ before observing  $\hat{\bm{\strain}}_{\text{obs}}(t_{i-1})$ and let $\mu^a(t_i)$ and $\Sigma^a(t_I)$ denote the mean and covariance of $v(t_i)$ give \textit{all previous} observations $\hat{\bm{\strain}}_{\text{obs}}(t_1), \ldots, \hat{\bm{\strain}}_{\text{obs}}(t_i)$.   Note that $\mu(t_i)$ and $\Sigma(t_i)$ are often called the forecast mean and covariance, while $\mu^a(t_i)$ and $\Sigma^a(t_i)$ are typically called the analyzed mean and covariance. 

Using the discrete Markov process in \eqref{eq:markov1}--\eqref{eq:markov2}, we can define the relationship between the analyzed mean and covariance at time $t_{i-1}$ and the forecast mean and covariance at time $t_i$ to be
\begin{eqnarray}
\mu(t_i)  & = & \bar{F} \mu^a(t_{i-1}) \label{eq:forecastMu} \\
\Sigma(t_i) & = & \bar{F} \Sigma^a(t_{i-1}) \bar{F}^T + \bar{Q}_{i-1} \label{eq:forecastCov}.
\end{eqnarray}

We can also exploit the Gaussianity of $v(t_i, \omega)$ to analytically derive the form of the analyzed mean $\mu^a(t_i)$ and covariance $\Sigma(t_i)$.
\begin{eqnarray}
\mu^a(t_i) &=& \mu(t_i) + K_i\left[ \hat{\bm{\strain}}_{\text{obs}}(t_i) -  \mu_{\text{obs}}(t_i) - H \mu(t_i)  \right] \label{eq:analyzedMean}\\
\Sigma^a(t_i) &=&\Sigma(t_i) - \Sigma(t_i) H^T \left[H \Sigma(t_i) H^T + \Sigma_{\epsilon\epsilon} \right]^{-1} H \Sigma(t_i) \label{eq:analyzedCov}
 \end{eqnarray}
where $K_i$ is the familiar Kalman gain
\begin{equation}
K_i = \Sigma(t_i) H^T \left[H \Sigma(t_i) H^T + \Sigma_{\epsilon\epsilon} \right]^{-1}.
\end{equation}
 
\paragraph{Smoothing} Kalman filtering, as in \eqref{eq:analyzedMean} and \eqref{eq:analyzedCov}, provides a way for estimating the state $v(t_i)$ given all the observations that happened on or before time $t_i$.   However, additional information is obtained with observations after $t_i$ and we would like to incorporate that into our characterization of $v(t_i)$ as well.  This problem, referred to as smoothing, aims to characterize the density $\pi( v(t_i) | \hat{\bm{\strain}}_{\text{obs}}(t_1), \ldots, \hat{\bm{\strain}}_{\text{obs}}(t_M) )$ for all $i$.  

The solution of the smoothing problem will be an update to the filtering distributions $\pi( v(t_i) | \hat{\bm{\strain}}_{\text{obs}}(t_1), \ldots, \hat{\bm{\strain}}_{\text{obs}}(t_i) )$.  To see this, consider the Markov properties of the model in \eqref{eq:markov1} and \eqref{eq:markov2}, which imply that the state $v(t_i)$ does not depend on directly on observations after $t_{i+1}$ if $v(t_{i+1})$.  More mathematically, we have
\begin{equation}
\pi( v(t_i) | \hat{\bm{\strain}}_{\text{obs}}(t_1), \ldots, \hat{\bm{\strain}}_{\text{obs}}(t_M) ) = \pi( v(t_i) | v(t_{i+1}), \hat{\bm{\strain}}_{\text{obs}}(t_1), \ldots, \hat{\bm{\strain}}_{\text{obs}}(t_i) ).
\end{equation}
Fortunately, this distribution can be characterized analytically in the linear-Gaussian setting.

Let $\mu^s(t_{i})$ and $\Sigma^s(t_{i})$ denote the mean and covariance of the solution to the smoothing problem at step $i$, i.e., $\pi( v(t_{i}, \omega) | \hat{\bm{\strain}}_{\text{obs}}(t_1), \ldots, \hat{\bm{\strain}}_{\text{obs}}(t_M) )$.  Notice that the smoothing solution is the same as the filtering solution at time $t_M$, so that $\mu^s(t_M) = \mu^a(t_m)$ and $\Sigma^s(t_M) = \Sigma^a(t_M)$.  For the remainder of the times, we can define the smoothing solution recursively, resulting in 
\begin{eqnarray}
\mu^s(t_i) &=& \mu^a(t_i) + C_i \left( \mu^s(t_{i+1})  - \mu(t_{i+1}) \right) \\
\Sigma^s(t_i) & = & C_i \left[\Sigma^s(t_{i+1}) - \Sigma(t_{i+1}) \right]  C_i^T
\end{eqnarray}
where $C_i = \Sigma^a(t_i) \bar{F}^T_{i+1} \Sigma(t_{i+1})^{-1}$.  These expressions for the well known Rauch-Tung-Striebel \cite{Rauch1965} fixed interval Kalman smoother.

\end{document}